\documentclass[DIV=calc, paper=a4, fontsize=11pt, twocolumn]{scrartcl}	 
\usepackage{authblk}

\usepackage{lipsum} 
\usepackage[english]{babel} 
\usepackage[protrusion=true,expansion=true]{microtype} 
\usepackage{amsmath,amsfonts,amsthm} 
\usepackage[svgnames]{xcolor} 
\usepackage[hang, small,labelfont=bf,up,textfont=it,up]{caption} 
\usepackage{booktabs} 
\usepackage{fix-cm}	 
\usepackage{graphicx}
\usepackage{hyperref}
\graphicspath{{figures/}}

\usepackage{sectsty} 
\allsectionsfont{\usefont{OT1}{phv}{b}{n}} 

\usepackage{fancyhdr} 
\usepackage{lastpage} 

\usepackage[utf8]{inputenc}

\usepackage[switch]{lineno}

\title{The diffusion metrics of African Swine Fever in Wild Boar}
\author{Hartmut H. K. Lentz\footnote{hartmut.lentz@fli.de} , Hannes~Bergmann, Carola~Sauter-Louis\\
\emph{Friedrich-Loeffler-Institut, Federal Research Institute for Animal Health, Institute of Epidemiology, 17493 Greifswald - Insel Riems}} 

\date{} 

\begin{document}
\maketitle 


\textbf{Abstract.}
To control African swine fever (ASF) efficiently, easily interpretable metrics of the outbreak dynamics are needed to plan and adapt the required measures.
We found that the spread pattern of African Swine Fever cases in wild boar follows the mechanics of a diffusion process, at least in the early phase, for the cases that occurred in Germany.
Following incursion into a previously unaffected area, infection disseminates locally within a naive and abundant wild boar population.
Using real case data for Germany, we derive statistics about the time differences and distances between consecutive case reports.
With the use of these statistics, we generate an ensemble of random walkers (continuous time random walks, CTRW) that resemble the properties of the observed outbreak pattern as one possible realization of all possible disease dissemination patterns.
The trained random walker ensemble yields the diffusion constant, the affected area, and the outbreak velocity of early ASF spread in wild boar.
These quantities are easy to interpret, robust, and may be generalized or adapted to different regions.
Therefore, diffusion metrics can be useful descriptors of early disease dynamics and help facilitate efficient control of African Swine Fever.

\par\vskip\baselineskip\noindent


\section{Introduction}
African swine fever virus (ASFV) causes an internationally spreading haemorrhagic pig disease with a massive socio-economic impact \cite{dixon2019african, Dixon2020}.
The current African swine fever (ASF) pandemic originated from disease incursion of genotype II ASFV in Georgia during 2007 \cite{asfEuropeReview2021}.
From there, ASF spread northwards into the Caucasus region, then further disseminated westwards into Europe, eastwards into Southeast Asia \cite{Dixon2020}, and even jumped across the Atlantic to threaten the Americas with outbreaks reported in the Dominican Republic and Haiti in 2021 \cite{oie-wahis}.
Since the start of the pandemic, an estimated quarter of the global domestic pig population has been decimated by the disease, causing food insecurity and economic losses on an unprecedented global scale \cite{you2021chinaeconomics,10.3389/fvets.2021.686038,10.3389/fvets.2020.00634}.
Particularly during the early phases following new ASF incursion, well informed anticipation of disease spread is critical for controlling the disease efficiently.

As a consequence of the incursion into Georgia in 2007, ASF (genotype II) reached the territory of the European Union (EU) in 2014, when first ASF cases were reported in wild boar in Lithuania and Poland \cite{Woniakowski2016, Pautienius, Maciulskis2020}.
Since then, and despite ongoing control efforts as well as intensive study of disease dynamics, ASF has been moving predominantly in western direction, affecting many more EU countries \cite{EFSA2020}.
Among them, in 2020, ASF has also reached Germany \cite{SauterLouisASFGermany}, where the disease continues to spread in initially distinct spatial clusters \cite{SauterLouis2021GermanClusters}.

In eastern and central Europe, wild boar seem to represent the predominant, disease-sustaining reservoir host in the current European ASF scenario.
This is based on the spatial extent of cases in this pig type \cite{Bergmann2021}, as well as their critical role in disease transmission through persistence of virus in the environment \cite{carlson,Chenais2018,Dixon2020,Mazur-Panasiuk2019}.
Infected wild boar that succumb to the disease, which is characterized by a case/fatality ratio of $> 90$ \%, may harbor infectious virus for weeks, if not months, after the death of the animals.

Unexpected occurrences of wild boar-ASF cases in locations that are a long distance away from the nearest previously affected area, such as suspected point incursions into the Czech Republic \cite{Cukor2020}, Belgium \cite{Linden2019}, into the western part of Poland \cite{Mazur-Panasiuk2019}, or into Northern Italy \cite{oie-wahis}, indicate that ASF can be relocated in association with human activities.
However, typically ASF spreads in a gradual manner through infections and dissemination of disease in wild boar at a local scale.
Whilst ASF outbreaks in domestic pigs appear to be manageable in most countries, the gradual disease spread in wild boar is very difficult to control and often persists \cite{Chenais2019,Gogin2013,Schulz2019}.
Based on historic ASF case reports, average disease spread velocities of approximately 1 to 1.5 km per month have been estimated \cite{EFSA2018,EFSA2020,Podgorski2018}.
Control measures that efficiently manage ASF dissemination following new incursions require risk-based allocation of limited resources and rely on disease spread predictions that are locally applicable to the acute outbreak situation in the field.

Most models for African swine fever in wild boar depend on a large number of parameters and assumptions (see \cite{hayes2021mechanistic} for a comprehensive overview).
Therefore, they suffer from high complexity and it is difficult to draw practical conclusions from these models.
For controlling African swine fever, simple and easily understandable metrics are needed, such as the following:
Given a new occurrence of ASF:
(1) What is the affected area for the next time?, 
(2) How far does the epidemic reach from the index case over time?, and
(3) What is the velocity of spread?

In order to answer these questions, we take a perspective different from most predictive models:
What if the underlying process is not relevant, and the outbreak points are just generated by a random-process?
If we understand this process, we can compute all the desired metrics that are described above.

Even though ASF dynamics seem to be complex in general, disease dissemination appears to follow a remarkably simple pattern when considered on a local scale.
In fact, the local dynamics of cases appear as growing areas of new cases emerging, and long-distance jumps are extreme events \cite{EFSA2020}.
On the one hand, long-distance jumps are extremely hard to predict as they are presumably caused by human activity \cite{asfEuropeReview2021}.
On the other hand, short distance spread is mainly caused by wild boar and understanding its dynamics is crucial for efficient counter measures.
Therefore, robust metrics are needed to quantify the dynamics of local ASF outbreaks.

In order to provide such metrics, we follow the idea to describe the epidemic as a pure diffusion process.
Logically, an epidemic is actually not a pure diffusion process -- i.e. a reaction-diffusion process -- but if an epidemic can be modelled in an accurate way by a diffusion process, this allows to interpret the results mathematically in a relatively simple fashion.

Diffusion is a macroscopic process that can be microscopically described as a stochastic process, also known as Brownian motion \cite{gardiner2009stochastic, metzler2000random,klafter2011first}.
In the context of ASF, the microscopic process is the single local outbreak, i.e. a spatial pattern of cases.
Once the logic of this microscopic process is understood and calibrated for the data, diffusion of disease spread can be extrapolated spatially and over time.

For a purely diffusive process, a similar approach has been used on human mobility data \cite{brockmann2006scaling}.
In the context of ASF, a probabilistic model considering random walks by wild boar with infection dynamics has been proposed previously \cite{taylor2021predicting}.
In contrast to that model, we consider the outbreak pattern generating process itself as a random walk.

Another model considers the diffusion around a primary case and including a habitat-suitability component \cite{korennoy2014spatio}.
The diffusion component in \cite{korennoy2014spatio}, however, is not time dependent and therefore the model is not suitable for temporal predictions.
A predictive model for ASF has been proposed in \cite{barongo2016mathematical}.
This approach is based on a compartment model and is therefore suitable for a prediction, even though it is mainly driven by assumptions instead of data which is typical for this model type.
Moreover, it does not consider a spatial component.

Besides mathematical models, individual based models have been used in order to estimate the transition parameters of ASF, based on real outbreak data \cite{lange2017elucidating}.
This model contains detailed data, and the movement and infection dynamics are considered explicitly.
In contrast to the model proposed in the present paper, however, the model in \cite{lange2017elucidating} requires a rather large number of assumptions.
Finally, the local wave front velocity of ASF has been modelled for Belgium in \cite{dellicour2020unravelling}.

All of the mentioned models provide good insights on the dynamics of ASF.
What has been missing so far is a model capturing the 'physics' behind the outbreak pattern.
As stated above, despite the fact that ASF is an infection process, it appears as a pure diffusion process on the map.
For this reason, we fit a diffusion model to the disease data in order to measure the diffusion parameters of the ASF epidemic directly.

\section{Material and Methods}
\subsection{Data}
\begin{figure}[htbp]
\includegraphics[width=\columnwidth]{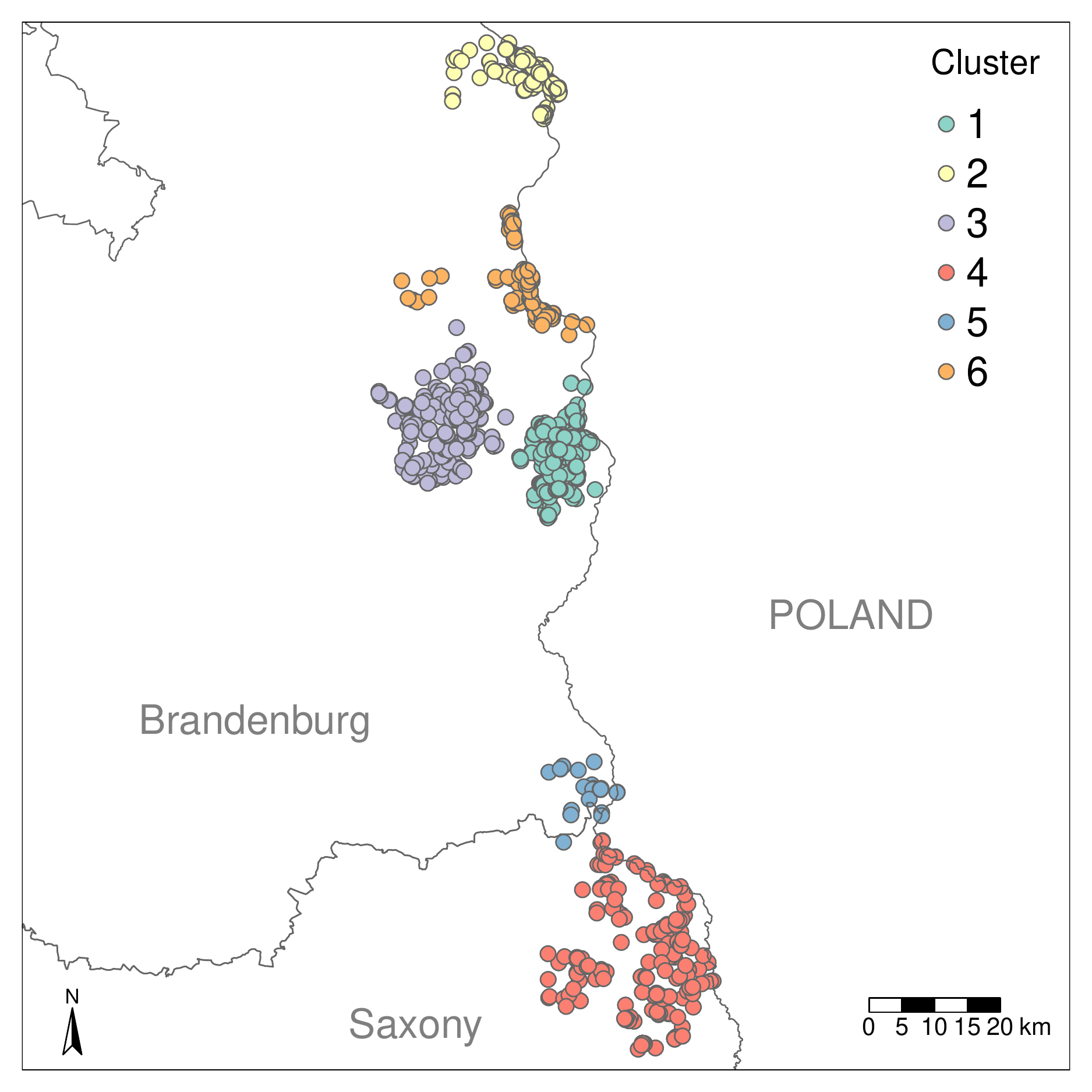}
\caption{African swine fever case data and its separation into clusters.}
\label{fig:data-clusters}
\end{figure}

We used the official ASF case data for Germany covering all cases from 10 September 2020 to 9 July 2021 from the national animal disease database (Tierseuchennachrichtensystem) \cite{kroschewski2006animal}.
Being in the early phase of the outbreak, we were able to separate the data into clusters \cite{SauterLouis2021GermanClusters}, as shown in Figure \ref{fig:data-clusters}.
In this paper, we will analyze one representative cluster in detail (Cluster 1), for reasons of simplicity.
All other clusters show similar microscopic patterns (see Supplementary Information) and we compare all clusters briefly in the results section. 

Each instance in the data set represents a \emph{case}, i.e. time and coordinates of a detected ASF-positive wild boar.
To distinguish between real data and data generated by our model, we use the term \emph{event} for the random walk model instead of \emph{case}.
For clarity, we refer to a cluster of ASF-cases in wild boar as an \emph{outbreak} (which should not be confused with occurrence of ASF in domestic pigs).

\begin{figure}[htbp]
\begin{center}
\includegraphics[width=\columnwidth]{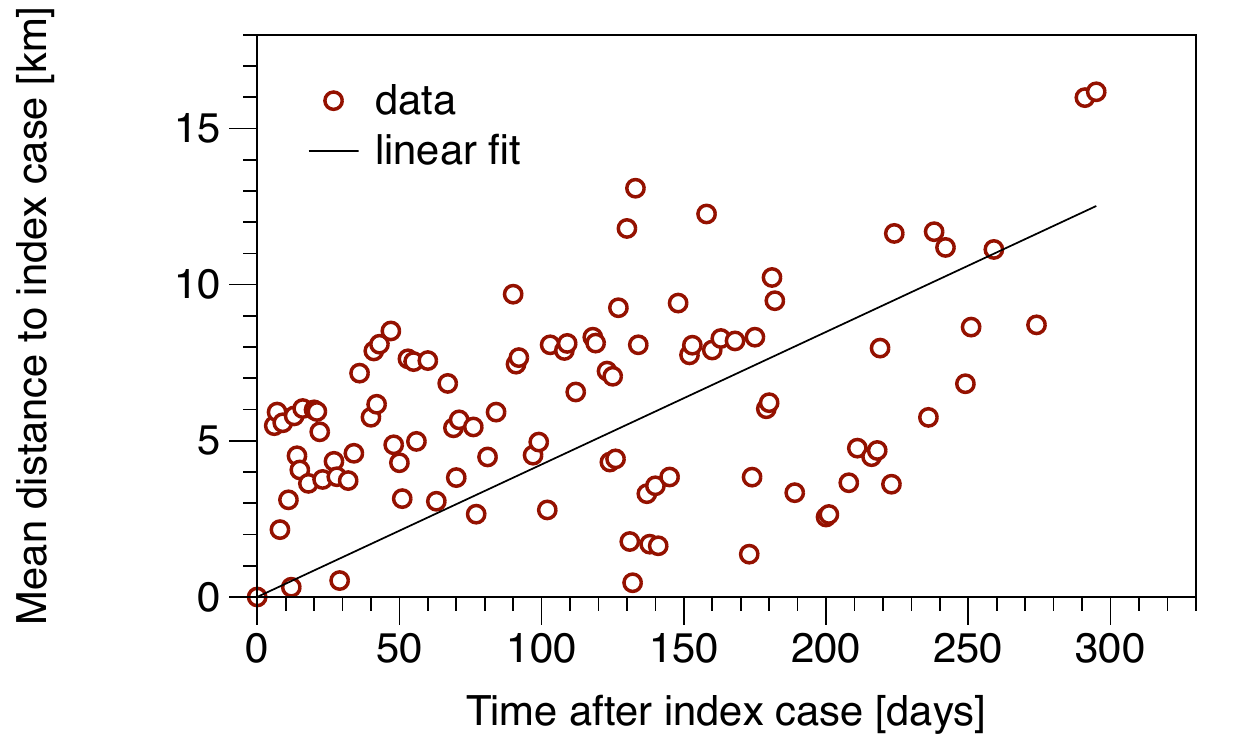}
\caption{Distance to index case as a function of time. Every data point represents one case. Where multiple cases were detected at one day, the y-axis presents the mean distance. Fitted slope is $0.042 \; \mathrm{km/day}$.}
\label{fig:App:disttoorigin}
\end{center}
\end{figure}

In order to get a first simple estimate of the outbreak velocity, we calculate the distance from each case to the index case over time.
This is shown in Figure \ref{fig:App:disttoorigin}.
Using a linear fit with vanishing intercept, we obtain a velocity of $0.042 \; \mathrm{km/day}$.
As shown below, this approach gives a good first estimate of the velocity, although it does not capture all features of the dynamics.

For a deeper understanding of the outbreak pattern, we consider the causal ordering of the cases in more detail.
It is important to stress that the data set itself does not contain any causal information between the cases.
Indeed, the measured data points represent an underlying -- and unknown -- infection tree that describes in detail, which case has caused which other case(s).

Since the exact relationships in this infection tree data are unknown, we estimate causality in the following way (a similar idea was used in \cite{EFSA2021epidemiological}):
\begin{enumerate}
\item Sort cases by time.
\item Generate a directed acyclic graph (DAG) $T=(V(t), E)$ with edge set $E = \emptyset$, where each node $v(t) \in V(t)$ is a case with time stamp $t$.
\item Connect nodes in $T$ with directed edges from case $s$ to case $t$ as follows: whenever the target case $t$ is after or at the same time as the source case $s$, draw a directed edge $(s, t)$.
Thus, the added edges $E \neq \emptyset$ in T comprise all possible causal connections between the cases.
\item Weight all edges with the reciprocal geographical distance between the respective nodes/cases.
(Vanishing distances are assigned a weight of zero.)
\item Finally, compute a minimum spanning tree on the now weighted DAG.
For this, we used the Chu-Liu/Edmond Algorithm \cite{chu1965shortest,edmonds1967optimum} implemented in \cite{hagberg2008exploring}.
\end{enumerate}

This procedure arranges the cases in a causally and geographically plausible order.
Using the minimum spanning tree, we obtain the distances between the cases, and subsequently the jump length distribution.

The empirical distribution of waiting times was directly derived from the outbreak data.
We sort the cases by time and compute the differences between consecutive cases yielding the waiting time distribution.

\subsection{Brownian Motion}
In the present work we consider the outbreak data as points that are seemingly generated randomly in space.
The only constraint is that new data points are generated in geographical and temporal closeness to existing points.
We hereby establish the simple assumption that new data points are somehow related to existing data points.
If we assume in the first place that every existing data point generates exactly one new data point at the next time step, the generating process would be Markovian.
On closer consideration, however, this assumption is not valid, since waiting times occur between the cases, and thus secondary outbreaks can be later in the future.

Random walk processes that consider waiting times are called \emph{continuous time random walks} (CTRW).
Such a process works as follows:
A random walker is initiated at time $t=0$ (time of the first case) at a location, say $(x_0, y_0) = (0, 0)$ (location of the first case).
Then it waits for a random time $\tau_1$ and makes a jump of random length $l_1$ in a random direction.
Thereafter, it waits for a random time $\tau_2$ and performs another jump of random length $l_{2}$ and so forth.
We assume that jump lengths and waiting times are uncorrelated.
Jump lengths are sampled from a distribution $\phi (l)$ and the waiting times from a distribution $\psi (\tau)$.
In this work, the distributions of $\phi (l)$ and $\psi (\tau)$ are determined from the outbreak data.
Hence, we generate synthetic outbreak data that is statistically equivalent to the observed data.

The CTRW is implemented as follows:
\begin{itemize}
\item Start at the coordinates of the index case. Set these $(x_0, y_0) = (0, 0)$.
\item Sample time jumps from the waiting-time distribution $\psi (\tau)$ and generate a sequence of time points (event points) following the time jumps. 
\item For each event point: sample a jump length from the jump-length distribution $\phi (l)$ and perform a step in a random direction.
\end{itemize}
The latest event determines the duration $T$ of the random walk.
We refer to one realization of a CTRW as a \emph{trajectory} $X(t)$.

It is important to stress the fact that, in contrast to an epidemic process, a random walker trajectory can only be at one location at a time.
In order to address this factor, we correct the available time for the random walker.

\subsection{Time correction in random walk}
Besides the above-mentioned waiting times, another crucial assumption for random walks does not hold for the outbreak data.
On the one hand, there can be multiple cases at every time step, that is, an epidemic can be at multiple locations at the same time.
On the other hand, a classic random walker can only be at one location at a time.
In order to resolve this issue, we use the following idea:
Let the random walk have a maximum duration of $T$.
In the easiest situation, exactly one case would occur at each day.
However, considering the situation where on average $M$ cases occur per day, the random walker must have the ability to generate these cases/events without spending time.
We call $M$ the \emph{multiplicity} of the process.
As an example, if we have $M=3$ cases per day and the maximum duration is $T=100$ days, then the random walker has $MT=300$ available days for generating these 300 events in total.
Finally, in order to return to the original time scale, we rescale the new maximum duration (300 days) back to the initial value (100 days).

\subsection{Diffusion Coefficient, expected radius, and velocity}
The Brownian motion described above is a single realization of a microscopic random process.
Averaging over a large number of random walks yields the macroscopic properties of the process.
Since every random walker can walk in a different direction, the expected location is $\left< X(t)\right> =(x_0, y_0)= 0$ for all times $t$ (the brackets $\left< \cdot \right>$ refer to the average over all random walkers).

For large times $t$ a single random walker is expected to be located at a great distance from the origin.
Therefore, the \emph{mean squared displacement} (MSD) $\left< X(t) ^2 \right>$ increases with time.
The detailed form of the MSD has to be determined empirically.
In case the MSD follows a linear relation, i.e. $\left< X(t) ^2 \right> \sim t$, the corresponding macroscopic process is called \emph{normal diffusion}.
In that case
\begin{equation}\label{eq:diffcoeff}
\left< X(t) ^2 \right> = 4 D t
\end{equation}
and the constant $D$ is the \emph{diffusion coefficient}.
Equation \eqref{eq:diffcoeff} represents the variance of the random walkers' positions after time $t$.

The square root of the MSD is the expected radius, on which all random walkers are expected to be after time $t$, i.e.
\begin{equation}\label{eq:radius}
r(t) = \sqrt{\left< X(t) ^2 \right>} = \sqrt{4Dt}.
\end{equation}
We identify this quantity with the radius of the affected area or the distance between the index case and the wave front.

Finally, the velocity of the wave front $v(t)$ can be defined as the change of the radius with respect to time, thus
\begin{equation}\label{eq:velocity}
v(t) = \frac{dr(t)}{dt} = \sqrt{ \frac{D}{t}}.
\end{equation}
Note that $r(t)$ and $v(t)$ are not linear.

Our implementation of the mentioned methods is available online \cite{github-random-walk}.

\section{Results}

\begin{figure}[htbp]
\includegraphics[width=\columnwidth]{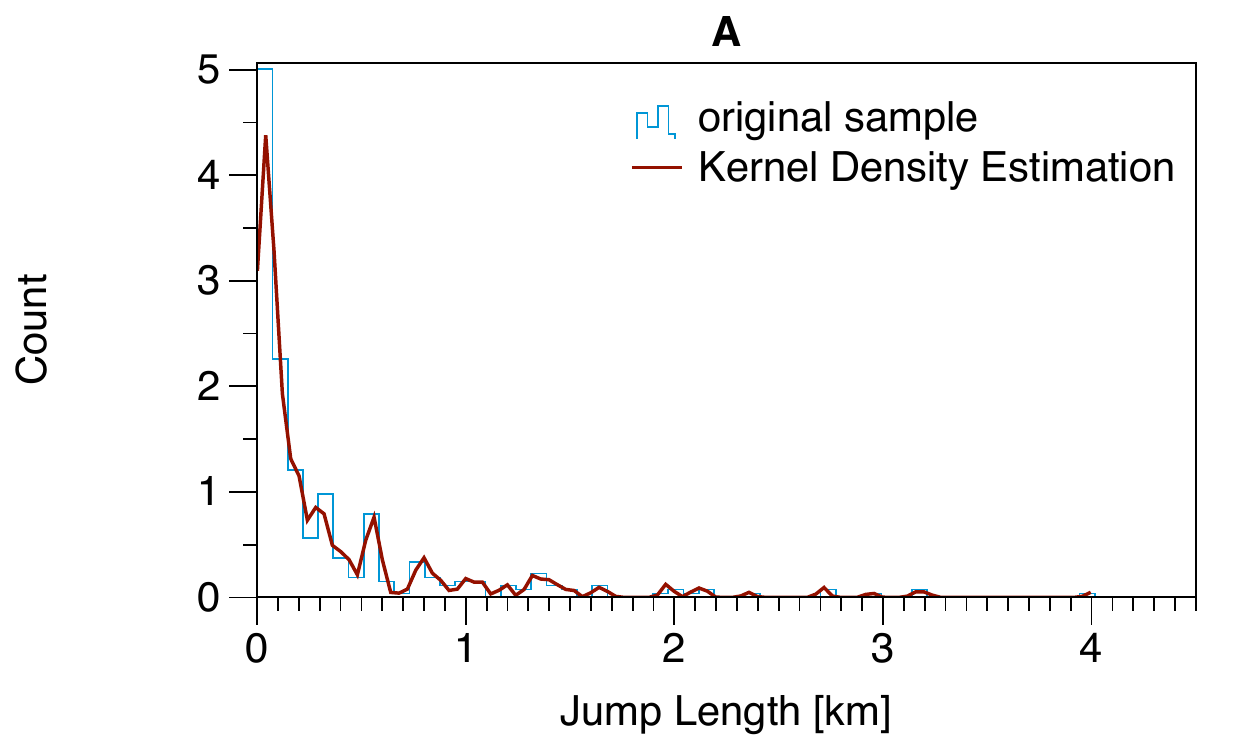}
\includegraphics[width=\columnwidth]{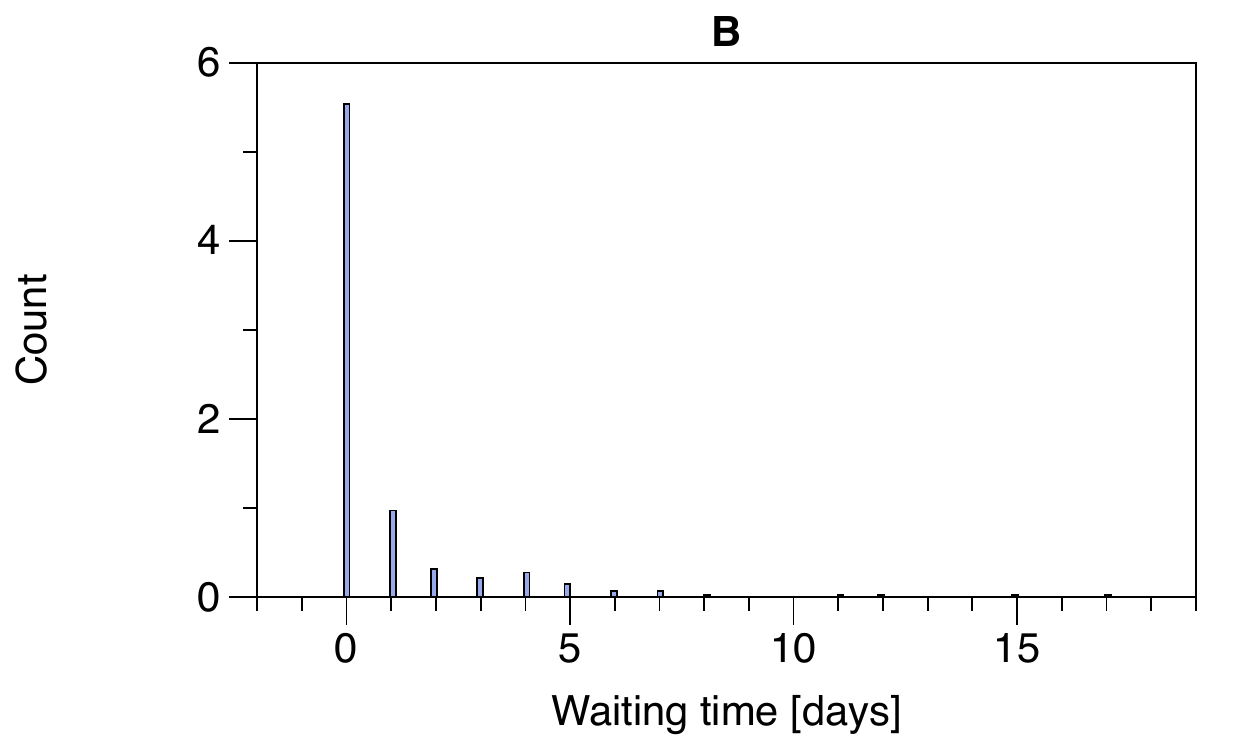}
\caption{Jump length distribution (A) and waiting time distribution (B) derived from the outbreak data of Cluster 1. (The jump length distribution is for cases ordered using the minimum spanning tree).}
\label{fig:jump-time-dist}
\end{figure}

To bring all cases in a plausible order, we first sort the outbreak data by detection time and compute the minimum spanning tree.
This tree provides us with the distribution of shortest jump lengths.
We then generate the waiting time distribution directly from the outbreak data.
Both distributions are shown in Figure \ref{fig:jump-time-dist}.

\begin{figure}[htbp]
\begin{center}
\includegraphics[width=\columnwidth]{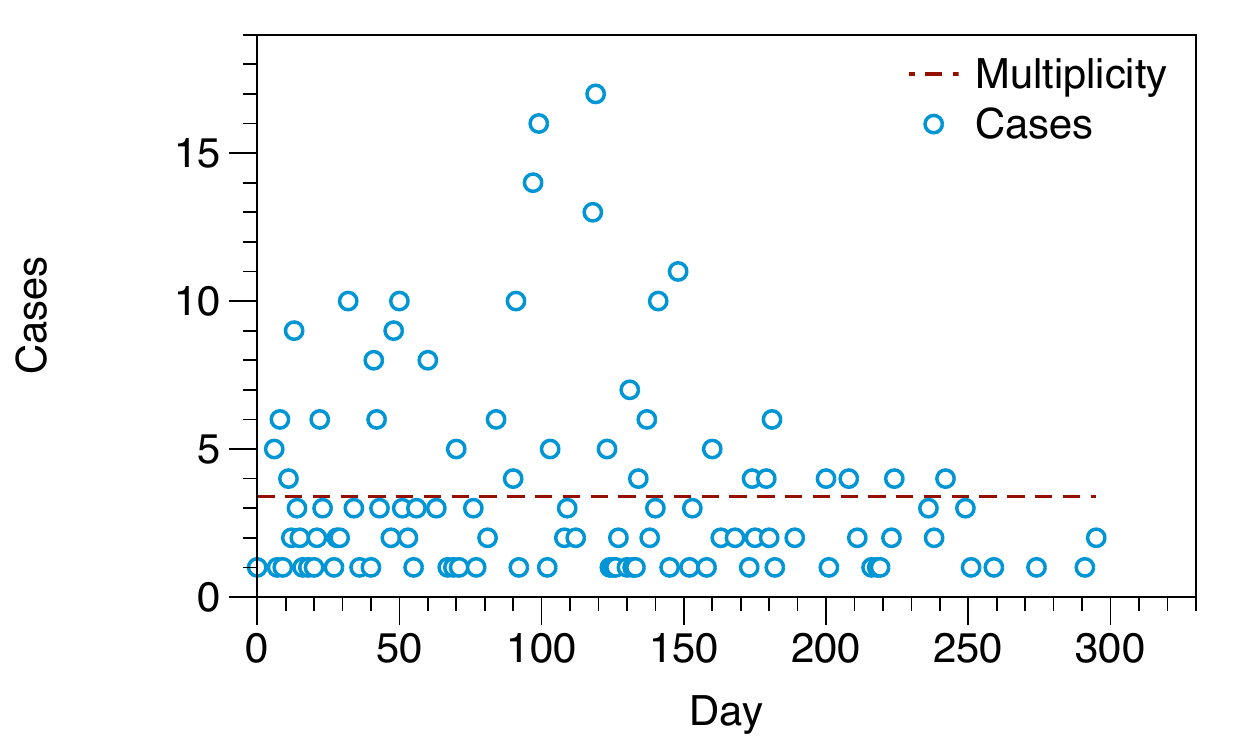}
\caption{Number of cases per day of cluster 1. The multiplicity is 3.4.}
\label{fig:eventsperday}
\end{center}
\end{figure}

As mentioned above, a random walker cannot be at multiple locations at the same time, as opposed to epidemic processes, where multiple cases can occur simultaneously.
This is exemplified by Cluster 1 in Figure \ref{fig:eventsperday}.

On average 3.4 cases occur per day over 300 days in total, i.e. the multiplicity of the process is $M=3.4$.
Therefore, we multiply the available time for the random walker by $M$.
This yields 1020 time steps which are afterwards rescaled to 300 days.

\begin{figure}[htbp]
\begin{center}
\includegraphics[width=\columnwidth]{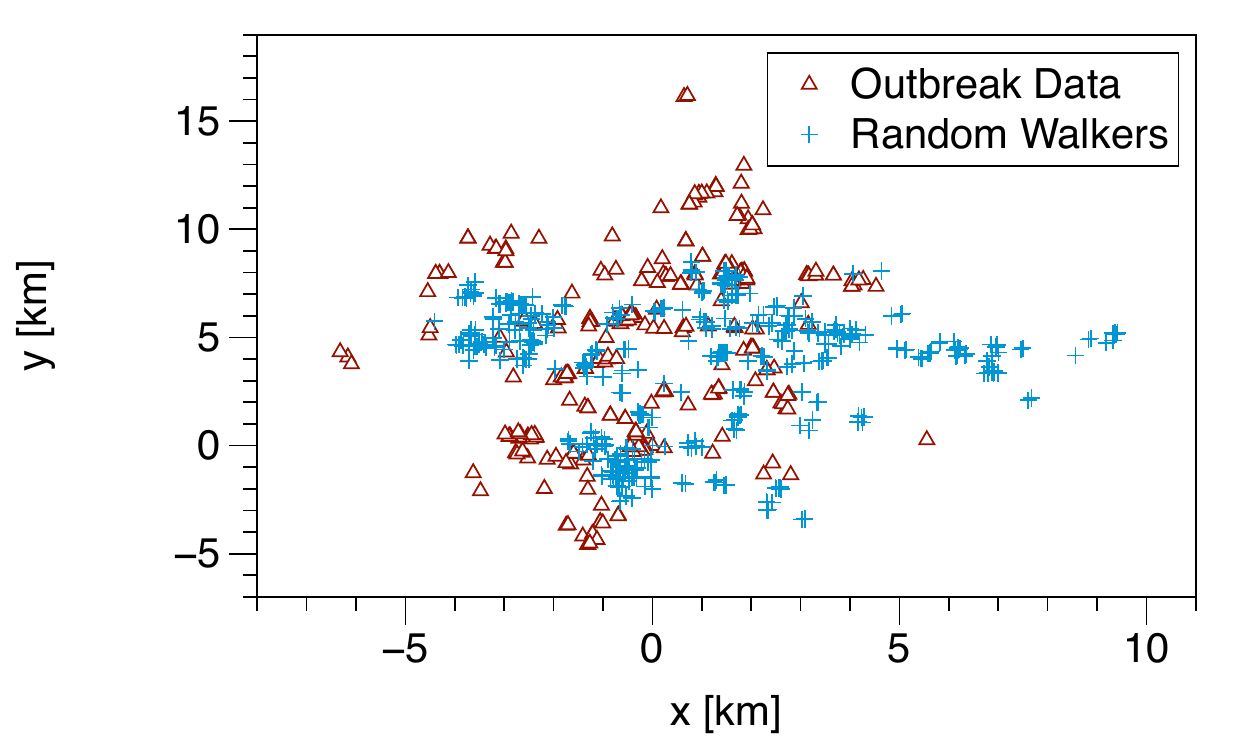}
\caption{Real outbreak data vs. one realization of a random walk for Cluster 1. The index case is set to coordinates (0, 0). Outbreak data from 300 days, random walk with multiplicity 3.4 resulting in 1020 steps that represent 300 days.}
\label{fig:datavsctrw}
\end{center}
\end{figure}

Using the distributions from Figure \ref{fig:jump-time-dist} and the multiplicity $M$ we generate an ensemble of 10,000 random walkers to guarantee statistical stability.
In order to get an impression of the microscopic properties of the random walks, we show one realization in Figure \ref{fig:datavsctrw}.
This realization appears to show considerable structural similarity to the observed outbreak data, in the sense that both sets of points appear to be sampled from a similar distribution.
Note that a random walk is an isotropic process, i.e. all directions are equally likely.
It is therefore on purpose that the outbreak data and the synthetic data points can be in different directions as long as they have a similar structure.

\begin{figure}[htbp]
\begin{center}
\includegraphics[width=\columnwidth]{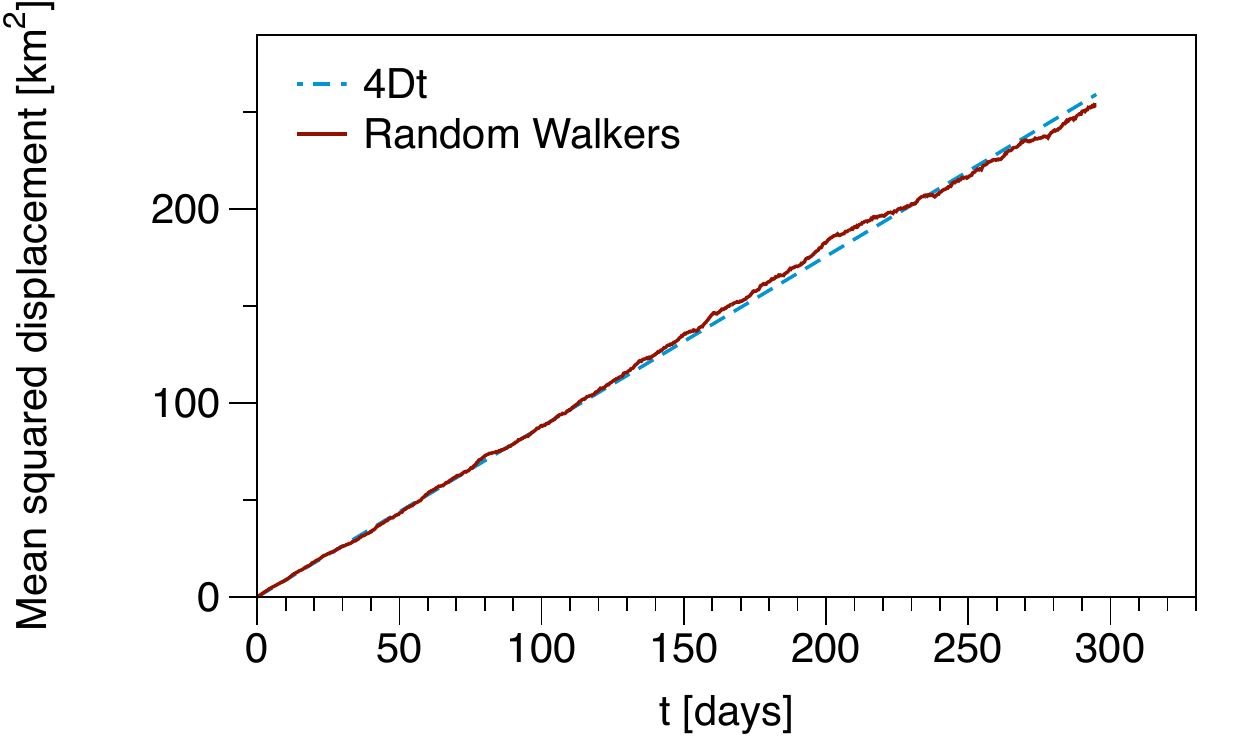}
\caption{Mean squared displacement for an ensemble of 10,000 random walkers (red line). The resulting diffusion constant $D$ follows from a linear fit (blue dashed line) which gives $D = (0.22 \pm 0.01) \; \mathrm{km}^{2} / \mathrm{day}$.}
\label{fig:MSD}
\end{center}
\end{figure}

We now study the macroscopic (diffusion) properties of the random walker ensemble.
Figure \ref{fig:MSD} shows the mean squared displacement (MSD) over an ensemble of 10,000 random walkers.
The MSD follows a linear form indicating that the measured distributions result in a normal diffusion process.
Using a linear fit, we obtain a diffusion coefficient of $D = (0.22 \pm 0.01) \; \mathrm{km}^{2} / \mathrm{day} $.
This value is a median over all realizations and the error is the inter-quartile range.

\begin{figure}[htbp]
\begin{center}
\includegraphics[width=\columnwidth]{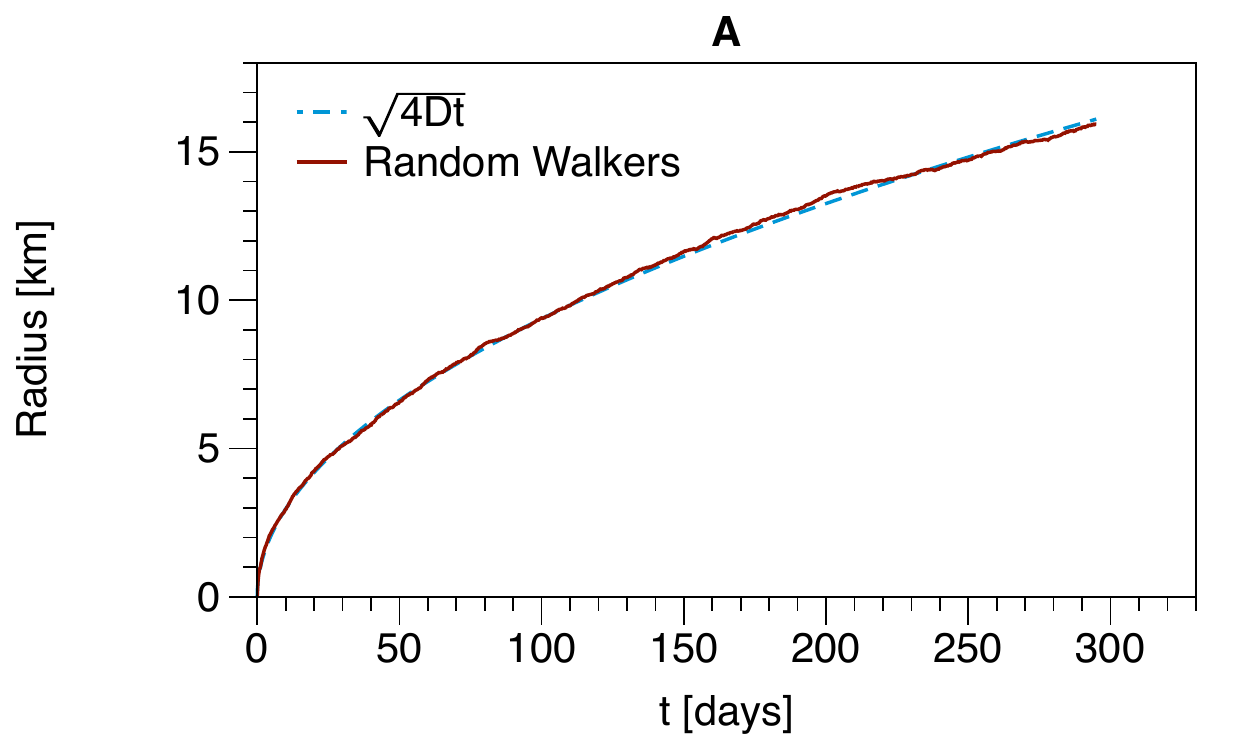}
\includegraphics[width=\columnwidth]{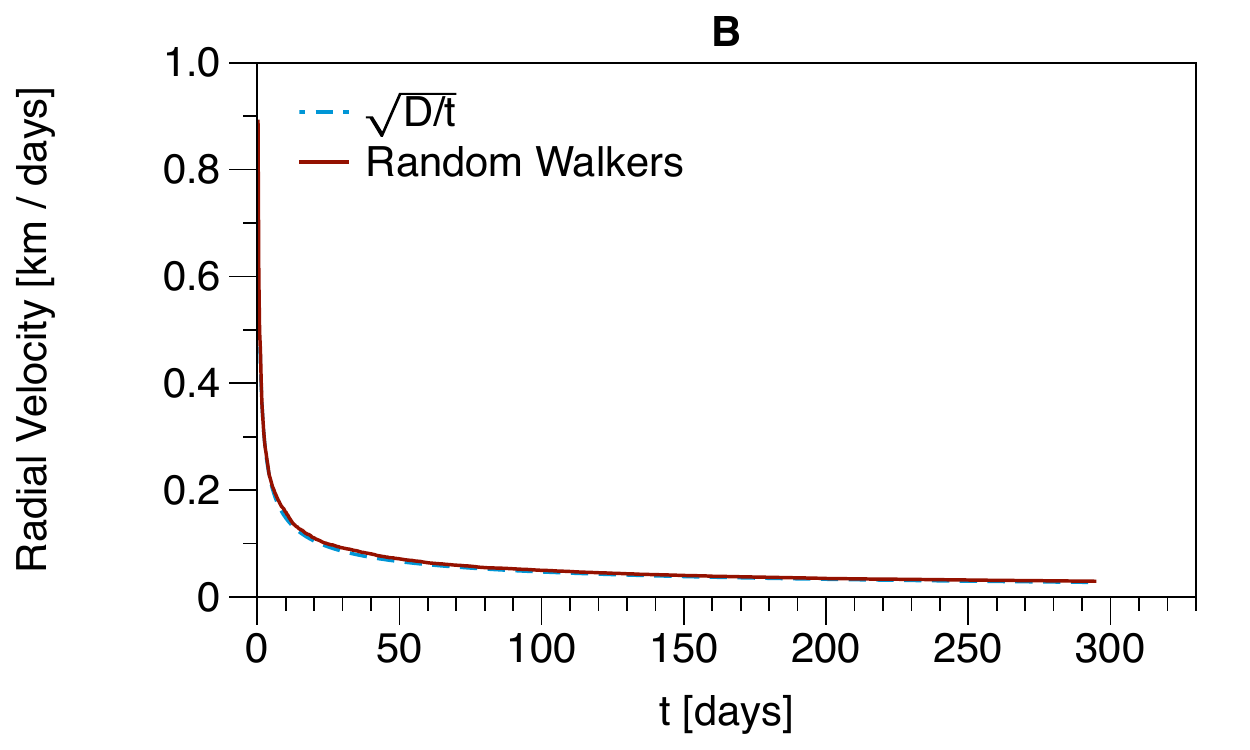}
\caption{A: Radius of the area where all random walkers are likely to be contained after time $t$. B: Radial velocity of the area growth. Red lines are mean values over the random walker ensemble, blue dashed lines are analytical, using the diffusion constant.}
\label{fig:radiusvelocity}
\end{center}
\end{figure}

The radius of the affected area follows the square root shaped relation shown in Figure \ref{fig:radiusvelocity} A.
After a steep increase in the early phase of the epidemic, the radius grows over time, but the front velocity decreases.
Note that the slowing down of the wave front cannot be captured by the simple linear approach used in Figure \ref{fig:App:disttoorigin}.
The wave front velocity over time is shown in Figure \ref{fig:radiusvelocity} B.
The latter shows a quasi-constant behavior in the time scale of interest, i.e. roughly 0.04 km/day measured 150 days after the first case.

\subsection{Comparison between the clusters}
\begin{figure}[htbp]
\begin{center}
\includegraphics[width=\columnwidth]{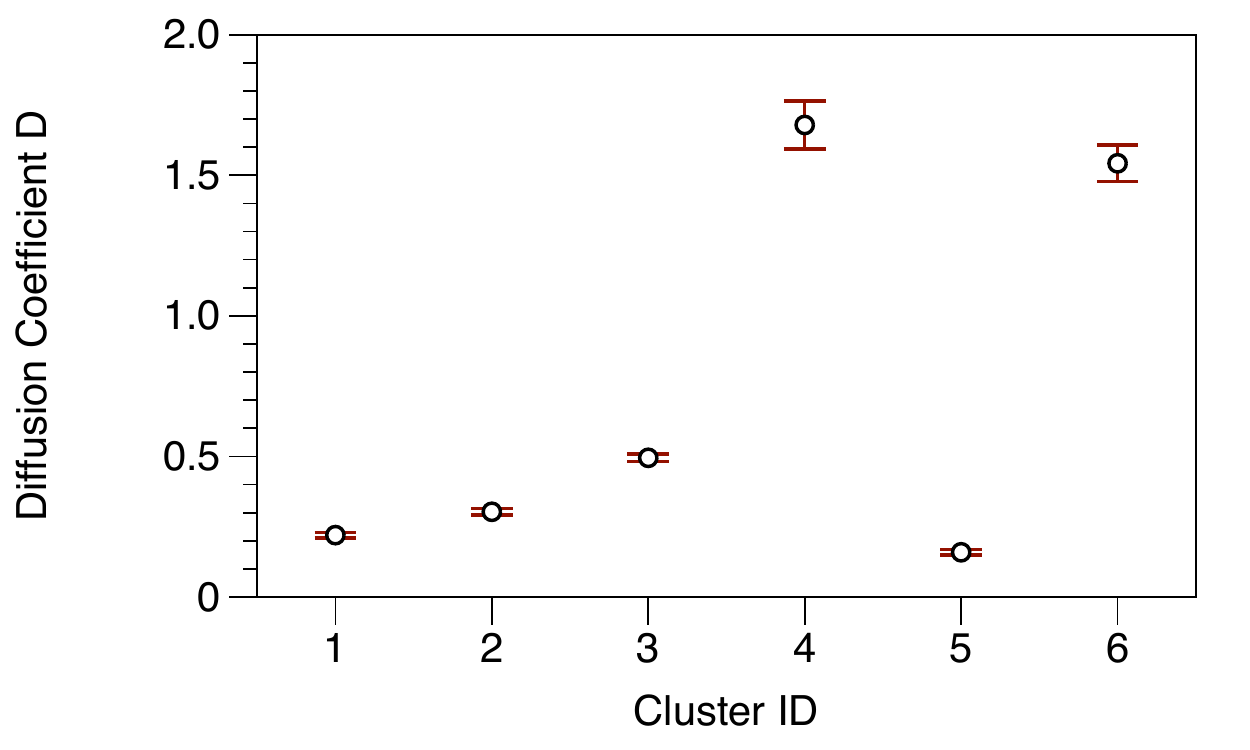}
\caption{Diffusion constants for all clusters. Error bars show the inter-quartile distance. Each data point is for 10,000 simulations.}
\label{fig:DvsClusterID}
\end{center}
\end{figure}

So far, we have only studied one selected cluster.
In Figure \ref{fig:DvsClusterID} we show the diffusion coefficients for all clusters.
Each value is a median over 10,000 simulations.
The error bars represent the inter quartile ranges.

The figure demonstrates that even if there are certain differences in the clusters, their diffusion coefficients show a remarkable similarity.
Most diffusion constants lie in a band between 0.2 and 0.5 $\mathrm{km}^{2} /\mathrm{day}$.

We provide a detailed description of the diffusion metrics for all clusters in the Supplementary Information.

\section{Discussion}
In the present study we have considered the outbreak propagation of African swine fever as a diffusion process.
Instead of making assumptions on wild boar movements, we focussed on the process generating the outbreak data directly.
Although this is an abstract concept, it allows us to measure the physical properties of the observed outbreak pattern.

Assuming that the outbreak propagation follows a random walk appears drastic, since in contrast to a random walk, new cases can appear at multiple locations at the same time.
This could be modelled by a branching process, where the random walker can reproduce itself.
As it has turned out, however, such a branching process becomes irrelevant, whenever all random walkers are not restricted in their possible location.
This property of the model is supported by the fact that infected animals can freely return to already infected areas, that is, the disease cannot be pushed out of already infected areas in the considered early phase of the outbreak.

Although our results provide simple metrics for the propagation of ASF, the computation of these metrics is not trivial in general.
On the one hand, estimating the wave front velocity using the simple linear distance to the index case has turned out to give a value remarkably similar to that of our model.
On the other hand, this simple approximation does not capture the slowing down of the wave front over time that is predicted by our model.
For the random walk model, besides the needed Monte-Carlo-simulations, finding the distributions for waiting times and jump lengths requires manual adjustments.
These could be optimized using a hyper-parameter-tuning scheme.
To obtain interpretable results, the real outbreak clusters should be not constrained, as it may be the case due to geographical barriers (rivers, roads, fences, etc.).
The more the real outbreak clusters are constrained, the more manual adjustment is required.
Considering the time data for the cases, the random walk approach has proven useful, although we have used the date, when ASF infection in dead wild boar were confirmed -- and not the date, when the animals died. 
In a next step, the estimated death times of the wild boar could be used, by applying the minimal post-mortem-interval \cite{probst2020estimating}.

As we have demonstrated in Figure \ref{fig:DvsClusterID}, the properties of some of the clusters are remarkably similar.
This seems to be reasonable, as the counter measures implemented overall are similar in all of the clusters.

Nevertheless, Clusters 4 and 6 show higher diffusion coefficients.
In the case of Cluster 4, this could be due to the fact that the time needed for fences to be erected was longer than in other cluster areas. 
Moreover, the first cases occurred along an extended area of the border without any expansion for the first 80 days. 
For Cluster 6, this could be caused by the fact that the disease occurred in an urban area, which did not allow for implementation of the same control measures as in the other clusters.
Moreover, the different diffusion coefficient might be caused by the fact that the cases occurred along an extended area at the German-Polish border, thus showing a high degree of constraint (see Figure \ref{fig:data-clusters}, and Supplementary Information for more details)
It is important to stress the fact that this constraint is caused primarily by the data availability and not by the underlying process.
That is, we would expect to get a more consistent picture here, if Polish data would have been included in the analysis.

Consequently, we state that the observed patterns follow a general mechanism, at least for this data set representing a particular area in Germany.
In conclusion, it seems possible to derive a general diffusion law for this kind of setting, which might be helpful for disease control.

\section{Acknowledgement}
This project has received funding from the European Union’s Horizon 2020 research and innovation programme under grant agreement No 773701.

\section{Author contributions}
H.L. designed the study and conducted the computations.
H.L., H.B., and C.S. wrote the manuscript.
C.S. contributed the data.

\section{Competing interests}
The authors declare no competing interests.

\section{Data availability}
The dataset used in this study can be made available for researchers upon request.

\section{Code availability}
The code is available online on https://github.com/hartmutlentz/RandomWalker2D.


\bibliographystyle{unsrt}

\begin{thebibliography}{10}

\bibitem{dixon2019african}
LK~Dixon, H~Sun, and H~Roberts.
\newblock African swine fever.
\newblock {\em Antiviral research}, 165:34--41, 2019.

\bibitem{Dixon2020}
L.~K. Dixon, K.~Stahl, F.~Jori, L.~Vial, and D.~U. Pfeiffer.
\newblock African swine fever epidemiology and control.
\newblock {\em Annu Rev Anim Biosci}, 8:221--246, 2020.

\bibitem{asfEuropeReview2021}
Carola Sauter-Louis, Franz~J Conraths, Carolina Probst, Ulrike Blohm, Katja
  Schulz, Julia Sehl, Melina Fischer, Jan~Hendrik Forth, Laura Zani, Klaus
  Depner, Thomas~C Mettenleiter, Martin Beer, and Sandra Blome.
\newblock African swine fever in wild boar in europe-a review.
\newblock {\em Viruses}, 13(9):1717, 08 2021.

\bibitem{oie-wahis}
World~Organisation for Animal Health~(OIE).
\newblock World animal health information system (wahis), 2021.

\bibitem{you2021chinaeconomics}
Shibing You, Tingyi Liu, Miao Zhang, Xue Zhao, Yizhe Dong, Bi~Wu, Yanzhen Wang,
  Juan Li, Xinjie Wei, and Baofeng Shi.
\newblock African swine fever outbreaks in china led to gross domestic product
  and economic losses.
\newblock {\em Nature Food}, 2(10):802--808, 2021.

\bibitem{10.3389/fvets.2021.686038}
Thinh Nguyen-Thi, Linh Pham-Thi-Ngoc, Que Nguyen-Ngoc, Sinh Dang-Xuan, Hu~Suk
  Lee, Hung Nguyen-Viet, Pawin Padungtod, Thuy Nguyen-Thu, Thuy Nguyen-Thi,
  Thang Tran-Cong, and Karl~M. Rich.
\newblock An assessment of the economic impacts of the 2019 african swine fever
  outbreaks in vietnam.
\newblock {\em Frontiers in Veterinary Science}, 8, 2021.

\bibitem{10.3389/fvets.2020.00634}
Jarkko~K. Niemi.
\newblock Impacts of african swine fever on pigmeat markets in europe.
\newblock {\em Frontiers in Veterinary Science}, 7, 2020.

\bibitem{Woniakowski2016}
Grzegorz Wo{\'z}niakowski, Edyta Kozak, Andrzej Kowalczyk, Magdalena {\L}yjak,
  Ma{\l}gorzata Pomorska-M{\'o}l, Krzysztof Niemczuk, and Zygmunt Pejsak.
\newblock Current status of african swine fever virus in a population of wild
  boar in eastern poland (2014-2015).
\newblock {\em Archives of Virology}, 161(1):189--195, 2016.

\bibitem{Pautienius}
A.~Pautienius, J.~Grigas, S.~Pileviciene, R.~Zagrabskaite, J.~Buitkuviene,
  G.~Pridotkas, R.~Stankevicius, Z.~Streimikyte, A.~Salomskas, D.~Zienius, and
  A.~Stankevicius.
\newblock Prevalence and spatiotemporal distribution of african swine fever in
  lithuania, 2014-2017.
\newblock {\em Virol J}, 15(1):177, 2018.

\bibitem{Maciulskis2020}
P.~Maciulskis, M.~Masiulis, G.~Pridotkas, J.~Buitkuviene, V.~Jurgelevicius,
  I.~Jaceviciene, R.~Zagrabskaite, L.~Zani, and S.~Pileviciene.
\newblock The african swine fever epidemic in wild boar (sus scrofa) in
  lithuania (2014-2018).
\newblock {\em Vet Sci}, 7(1), 2020.

\bibitem{EFSA2020}
EFSA.
\newblock Epidemiological analyses of african swine fever in the european union
  (november 2018 to october 2019).
\newblock {\em Efsa Journal}, 18(1), 2020.

\bibitem{SauterLouisASFGermany}
C.~Sauter-Louis, J.~H. Forth, C.~Probst, C.~Staubach, A.~Hlinak, A.~Rudovsky,
  D.~Holland, P.~Schlieben, M.~Goldner, J.~Schatz, S.~Bock, M.~Fischer,
  K.~Schulz, T.~Homeier-Bachmann, R.~Plagemann, U.~Klaass, R.~Marquart, T.~C.
  Mettenleiter, M.~Beer, F.~J. Conraths, and S.~Blome.
\newblock Joining the club: First detection of african swine fever in wild boar
  in germany.
\newblock {\em Transbound Emerg Dis}, 2020.

\bibitem{SauterLouis2021GermanClusters}
C.~Sauter-Louis, K.~Schulz, M.~Richter, C.~Staubach, T.~C. Mettenleiter, and
  F.~J. Conraths.
\newblock {{A}frican swine fever: {W}hy the situation in {G}ermany is not
  comparable to that in the {C}zech {R}epublic or {B}elgium}.
\newblock {\em Transbound Emerg Dis}, Jul 2021.

\bibitem{Bergmann2021}
H.~Bergmann, K.~Schulz, F.~J. Conraths, and C.~Sauter-Louis.
\newblock {{A} {R}eview of {E}nvironmental {R}isk {F}actors for {A}frican
  {S}wine {F}ever in {E}uropean {W}ild {B}oar}.
\newblock {\em Animals (Basel)}, 11(9), Sep 2021.

\bibitem{carlson}
J.~Carlson, M.~Fischer, L.~Zani, M.~Eschbaumer, W.~Fuchs, T.~Mettenleiter,
  M.~Beer, and S.~Blome.
\newblock Stability of african swine fever virus in soil and options to
  mitigate the potential transmission risk.
\newblock {\em Pathogens}, 9(11), 2020.

\bibitem{Chenais2018}
E.~Chenais, K.~Stahl, V.~Guberti, and K.~Depner.
\newblock Identification of wild boar-habitat epidemiologic cycle in african
  swine fever epizootic.
\newblock {\em Emerg Infect Dis}, 24(4):810--812, 2018.

\bibitem{Mazur-Panasiuk2019}
N.~Mazur-Panasiuk, J.~Zmudzki, and G.~Wozniakowski.
\newblock African swine fever virus - persistence in different environmental
  conditions and the possibility of its indirect transmission.
\newblock {\em J Vet Res}, 63(3):303--310, 2019.

\bibitem{Cukor2020}
J.~Cukor, R.~Linda, P.~Vaclavek, P.~Satran, K.~Mahlerova, Z.~Vacek, T.~Kunca,
  and F.~Havranek.
\newblock Wild boar deathbed choice in relation to asf: Are there any
  differences between positive and negative carcasses?
\newblock {\em Preventive Veterinary Medicine}, 177, 2020.

\bibitem{Linden2019}
A.~Linden, A.~Licoppe, R.~Volpe, J.~Paternostre, C.~Lesenfants, D.~Cassart,
  M.~Garigliany, M.~Tignon, T.~van~den Berg, D.~Desmecht, and A.~B. Cay.
\newblock Summer 2018: African swine fever virus hits north-western europe.
\newblock {\em Transboundary and Emerging Diseases}, 66(1):54--55, 2019.

\bibitem{Chenais2019}
E.~Chenais, K.~Depner, V.~Guberti, K.~Dietze, A.~Viltrop, and K.~Stahl.
\newblock Epidemiological considerations on african swine fever in europe
  2014-2018.
\newblock {\em Porcine Health Manag}, 5:6, 2019.

\bibitem{Gogin2013}
A.~Gogin, V.~Gerasimov, A.~Malogolovkin, and D.~Kolbasov.
\newblock {{A}frican swine fever in the {N}orth {C}aucasus region and the
  {R}ussian {F}ederation in years 2007-2012}.
\newblock {\em Virus Res}, 173(1):198--203, Apr 2013.

\bibitem{Schulz2019}
K.~Schulz, E.~Olsevskis, C.~Staubach, K.~Lamberga, M.~Serzants, S.~Cvetkova,
  F.~J. Conraths, and C.~Sauter-Louis.
\newblock Epidemiological evaluation of latvian control measures for african
  swine fever in wild boar on the basis of surveillance data.
\newblock {\em Sci Rep}, 9(1):4189, 2019.

\bibitem{EFSA2018}
EFSA.
\newblock Epidemiological analyses of african swine fever in the european union
  (november 2017 until november 2018).
\newblock {\em EFSA Journal}, (11), 2018.

\bibitem{Podgorski2018}
T.~Podgorski and K.~Smietanka.
\newblock Do wild boar movements drive the spread of african swine fever?
\newblock {\em Transbound Emerg Dis}, 65(6):1588--1596, 2018.

\bibitem{hayes2021mechanistic}
Brandon~H Hayes, Mathieu Andraud, Luis~G Salazar, Nicolas Rose, and
  Timoth{\'e}e Vergne.
\newblock Mechanistic modelling of african swine fever: A systematic review.
\newblock {\em Preventive Veterinary Medicine}, page 105358, 2021.

\bibitem{gardiner2009stochastic}
Crispin Gardiner.
\newblock {\em Stochastic methods}, volume~4.
\newblock Springer Berlin, 2009.

\bibitem{metzler2000random}
Ralf Metzler and Joseph Klafter.
\newblock The random walk's guide to anomalous diffusion: a fractional dynamics
  approach.
\newblock {\em Physics reports}, 339(1):1--77, 2000.

\bibitem{klafter2011first}
Joseph Klafter and Igor~M Sokolov.
\newblock {\em First steps in random walks: from tools to applications}.
\newblock Oxford University Press, 2011.

\bibitem{brockmann2006scaling}
Dirk Brockmann, Lars Hufnagel, and Theo Geisel.
\newblock The scaling laws of human travel.
\newblock {\em Nature}, 439(7075):462--465, 2006.

\bibitem{taylor2021predicting}
Rachel~A Taylor, Tomasz Podg{\'o}rski, Robin~RL Simons, Sophie Ip, Paul Gale,
  Louise~A Kelly, and Emma~L Snary.
\newblock Predicting spread and effective control measures for african swine
  fever---should we blame the boars?
\newblock {\em Transboundary and Emerging Diseases}, 68(2):397--416, 2021.

\bibitem{korennoy2014spatio}
FI~Korennoy, VM~Gulenkin, JB~Malone, CN~Mores, SA~Dudnikov, and MA~Stevenson.
\newblock Spatio-temporal modeling of the african swine fever epidemic in the
  russian federation, 2007--2012.
\newblock {\em Spatial and spatio-temporal epidemiology}, 11:135--141, 2014.

\bibitem{barongo2016mathematical}
Mike~B Barongo, Richard~P Bishop, Eric~M F{\`e}vre, Darryn~L Knobel, and Amos
  Ssematimba.
\newblock A mathematical model that simulates control options for african swine
  fever virus (asfv).
\newblock {\em PloS one}, 11(7):e0158658, 2016.

\bibitem{lange2017elucidating}
Martin Lange and Hans-Hermann Thulke.
\newblock Elucidating transmission parameters of african swine fever through
  wild boar carcasses by combining spatio-temporal notification data and
  agent-based modelling.
\newblock {\em Stochastic environmental research and risk assessment},
  31(2):379--391, 2017.

\bibitem{dellicour2020unravelling}
Simon Dellicour, Daniel Desmecht, Julien Paternostre, C{\'e}line Malengreaux,
  Alain Licoppe, Marius Gilbert, and Annick Linden.
\newblock Unravelling the dispersal dynamics and ecological drivers of the
  african swine fever outbreak in belgium.
\newblock {\em Journal of Applied Ecology}, 57(8):1619--1629, 2020.

\bibitem{kroschewski2006animal}
K~Kroschewski, M~Kramer, A~Micklich, C~Staubach, R~Carmanns, and FJ~Conraths.
\newblock Animal disease outbreak control: the use of crisis management tools.
\newblock {\em Revue Scientifique et Technique-Office International des
  Epizooties}, 25(1):211, 2006.

\bibitem{EFSA2021epidemiological}
European Food Safety~Authority (EFSA), Daniel Desmecht, Guillaume Gerbier,
  Christian Gort{\'a}zar~Schmidt, Vilija Grigaliuniene, Georgina Helyes, Maria
  Kantere, Daniela Korytarova, Annick Linden, Aleksandra Miteva, et~al.
\newblock Epidemiological analysis of african swine fever in the european union
  (september 2019 to august 2020).
\newblock {\em EFSA Journal}, 19(5):e06572, 2021.

\bibitem{chu1965shortest}
Yoeng-Jin Chu.
\newblock On the shortest arborescence of a directed graph.
\newblock {\em Scientia Sinica}, 14:1396--1400, 1965.

\bibitem{edmonds1967optimum}
Jack Edmonds.
\newblock Optimum branchings.
\newblock {\em Journal of Research of the National Bureau of Standards, B},
  71:233--240, 1967.

\bibitem{hagberg2008exploring}
Aric Hagberg, Pieter Swart, and Daniel S~Chult.
\newblock Exploring network structure, dynamics, and function using networkx.
\newblock Technical report, Los Alamos National Lab.(LANL), Los Alamos, NM
  (United States), 2008.

\bibitem{github-random-walk}
https://github.com/hartmutlentz\\/RandomWalker2D, 2021.

\bibitem{probst2020estimating}
Carolina Probst, J{\"o}rn Gethmann, Jens Amendt, Lena Lutz, Jens~Peter Teifke,
  and Franz~J Conraths.
\newblock Estimating the postmortem interval of wild boar carcasses.
\newblock {\em Veterinary sciences}, 7(1):6, 2020.

\end{thebibliography}

\label{letzteseite}

\clearpage
\pagenumbering{arabic}
\renewcommand*{\thepage}{A\arabic{page}}

\section{Supplementary Information}
\renewcommand\thefigure{A.\arabic{figure}} 
\setcounter{figure}{0}

\subsection{Wave front velocities}
\begin{figure}[htbp]
\begin{center}
\includegraphics[width=\columnwidth]{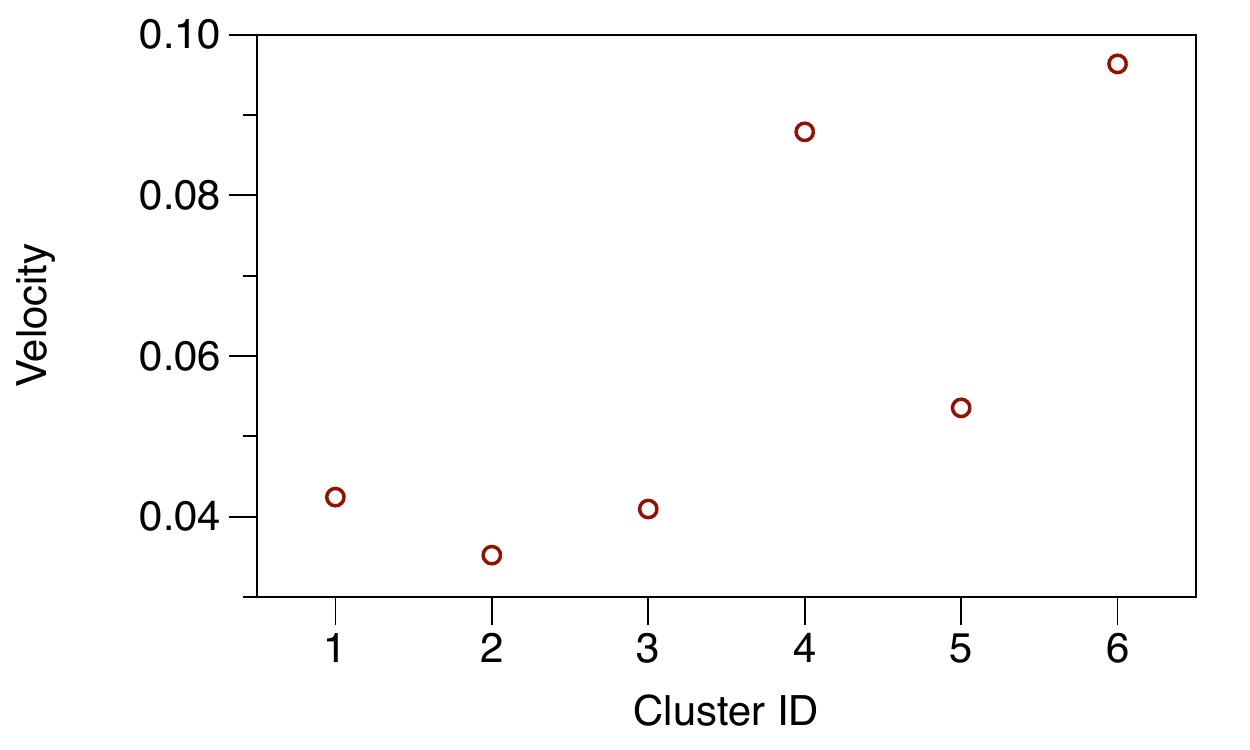}
\caption{Velocities for all clusters estimated using linear regression of the distances to the index case.}
\label{fig:velo_forall}
\end{center}
\end{figure}

We compare the estimated velocities found by simple linear regression as done in the main text (Figure \ref{fig:App:disttoorigin}).
The result is shown in Figure \ref{fig:velo_forall}.
As expected, the velocities are related to the diffusion coefficients for the considered clusters, i.e. they show similar values, except Clusters 4 and 6.

\subsection{Mean squared displacements of Clusters 2--6}

\subsubsection{Cluster 2}
\begin{figure}[htbp]
\begin{center}
\includegraphics[width=\columnwidth]{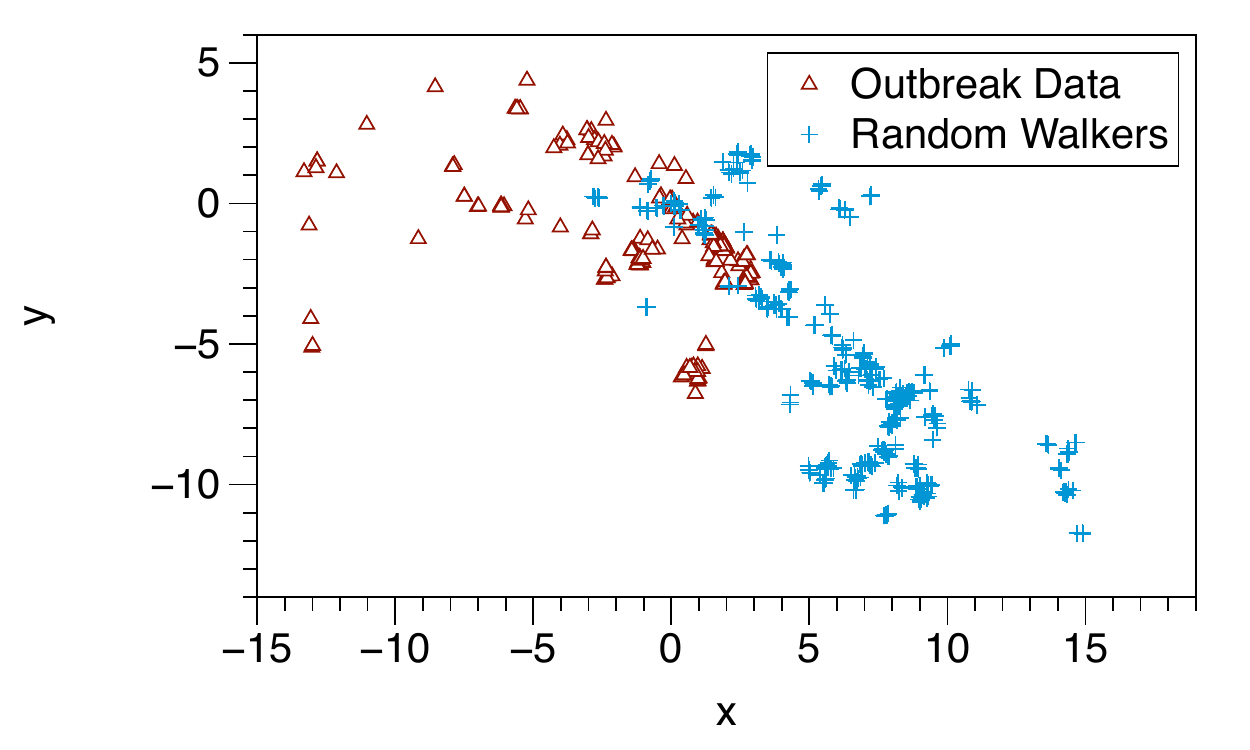}
\caption{Real outbreak data vs. one realization of a random walk for Cluster 2. The index case is set to coordinates (0, 0). Outbreak data from 265 days, random walk with multiplicity 4.1 resulting in 1087 steps.
}
\label{fig:clu2_realization}
\end{center}
\end{figure}

Figure \ref{fig:clu2_realization} shows a realization of a random walk trained on the outbreak data.
The multiplicity of the process is 4.1.

\begin{figure}[htbp]
\begin{center}
\includegraphics[width=\columnwidth]{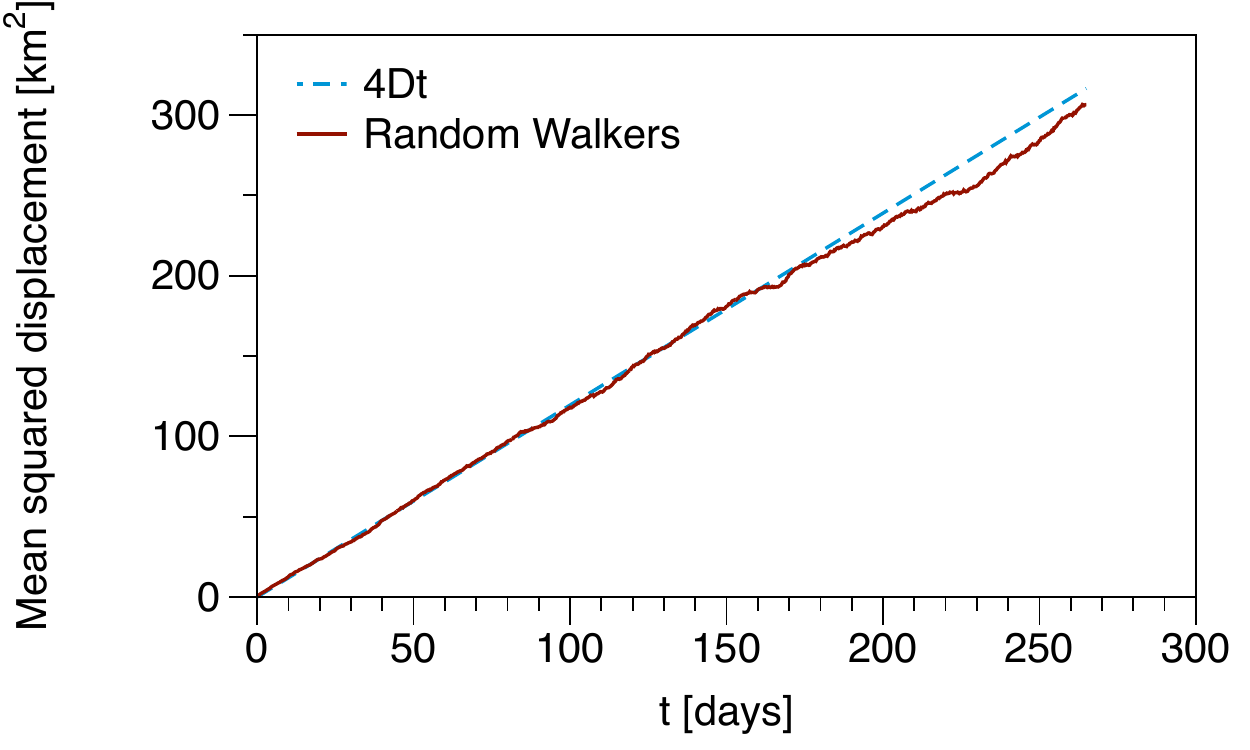}
\caption{Mean squared displacement for Cluster 2. Mean over 10,000 random walkers (red line). The resulting diffusion constant is $D = (0.30 \pm 0.01)\; km^2 / day$.}
\label{fig:clu2_msd}
\end{center}
\end{figure}

We show the mean squared displacement in Figure \ref{fig:clu2_msd}.
The diffusion coefficient is $D = (0.30 \pm 0.01) \; km^2 / day$.

\subsubsection{Cluster 3}
\begin{figure}[htbp]
\begin{center}
\includegraphics[width=\columnwidth]{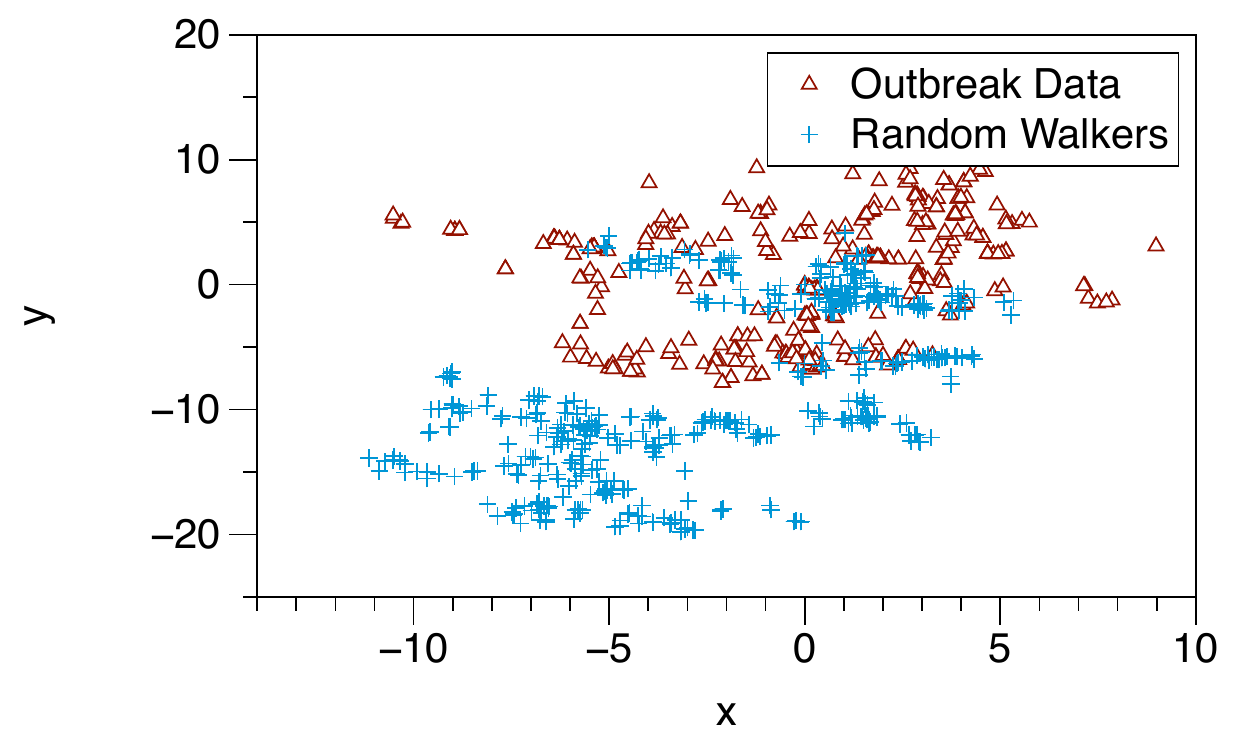}
\caption{Real outbreak data vs. one realization of a random walk for Cluster 3. The index case is set to coordinates (0, 0). Outbreak data from 245 days, random walk with multiplicity 3.6 resulting in 882 steps.
}
\label{fig:clu3_realization}
\end{center}
\end{figure}

Figure \ref{fig:clu3_realization} shows a realization of a random walk trained on the outbreak data.
The multiplicity of the process is 3.6.

\begin{figure}[htbp]
\begin{center}
\includegraphics[width=\columnwidth]{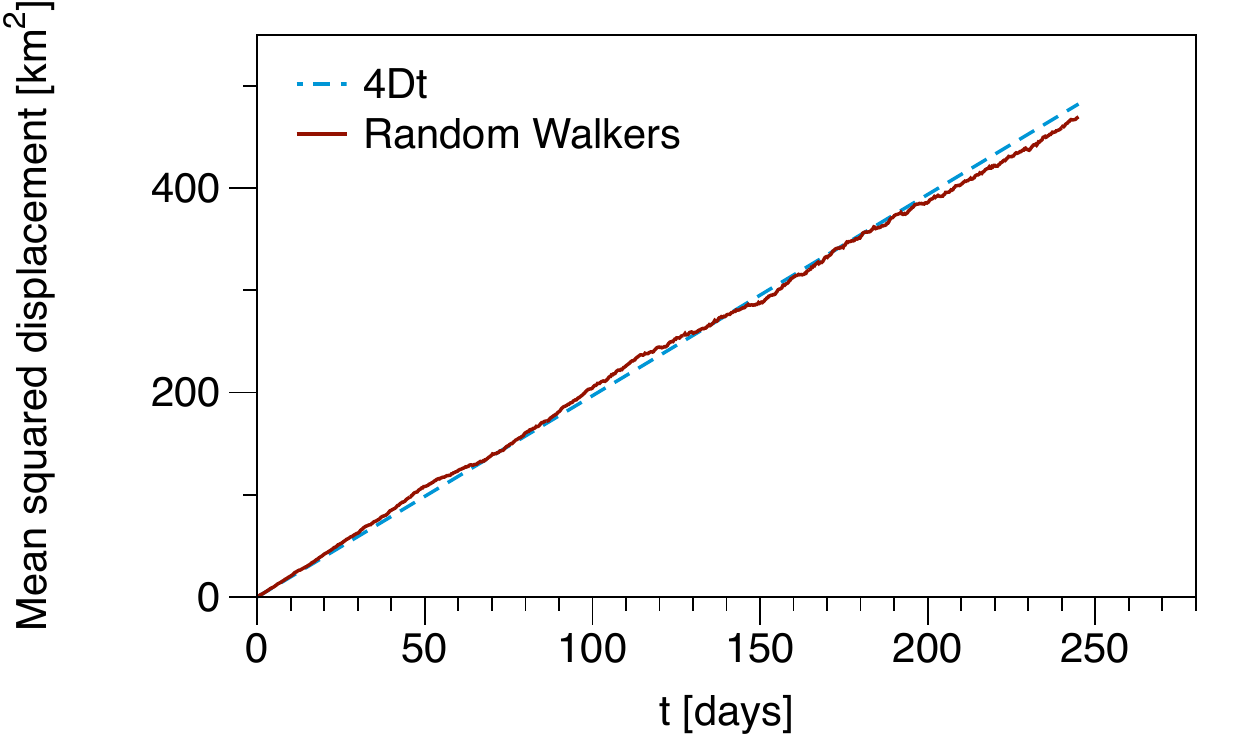}
\caption{Mean squared displacement for Cluster 3. Mean over 10,000 random walkers (red line). The resulting diffusion constant is $D = (0.50 \pm 0.01)\; km^2 / day$.}
\label{fig:clu3_msd}
\end{center}
\end{figure}

We show the mean squared displacement in Figure \ref{fig:clu3_msd}.
The diffusion coefficient is $D = (0.50 \pm 0.01) \; km^2 / day$.

\subsubsection{Cluster 4}
\begin{figure}[htbp]
\begin{center}
\includegraphics[width=\columnwidth]{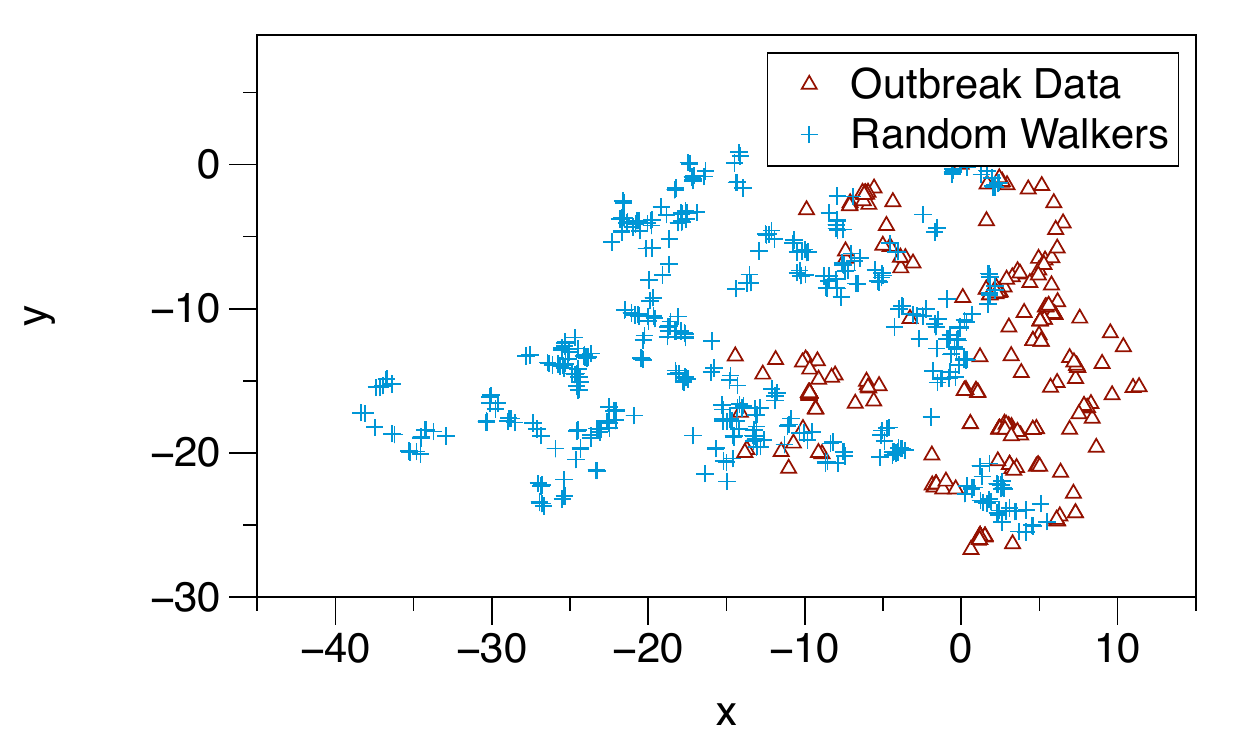}
\caption{Real outbreak data vs. one realization of a random walk for Cluster 4. The index case is set to coordinates (0, 0). Outbreak data from 249 days, random walk with multiplicity 4.7 resulting in 1170 steps.
}
\label{fig:clu4_realization}
\end{center}
\end{figure}

Figure \ref{fig:clu4_realization} shows a realization of a random walk trained on the outbreak data.
The multiplicity of the process is 4.7.

\begin{figure}[htbp]
\begin{center}
\includegraphics[width=\columnwidth]{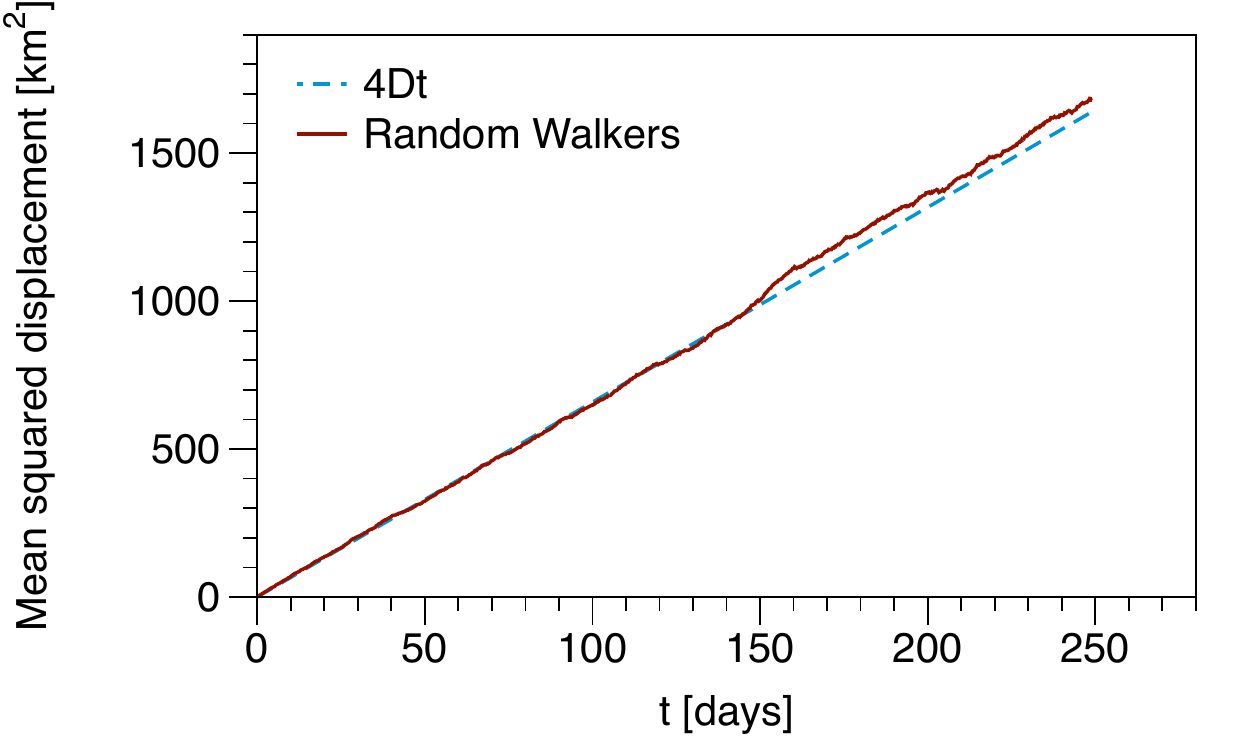}
\caption{Mean squared displacement for Cluster 4. Mean over 10,000 random walkers (red line). The resulting diffusion constant is $D = (1.69 \pm 0.09)\; km^2 / day$.}
\label{fig:clu4_msd}
\end{center}
\end{figure}

We show the mean squared displacement in Figure \ref{fig:clu4_msd}.
The diffusion coefficient is $D = (1.69 \pm 0.09) \; km^2 / day$.

\subsubsection{Cluster 5}

\begin{figure}[htbp]
\begin{center}
\includegraphics[width=\columnwidth]{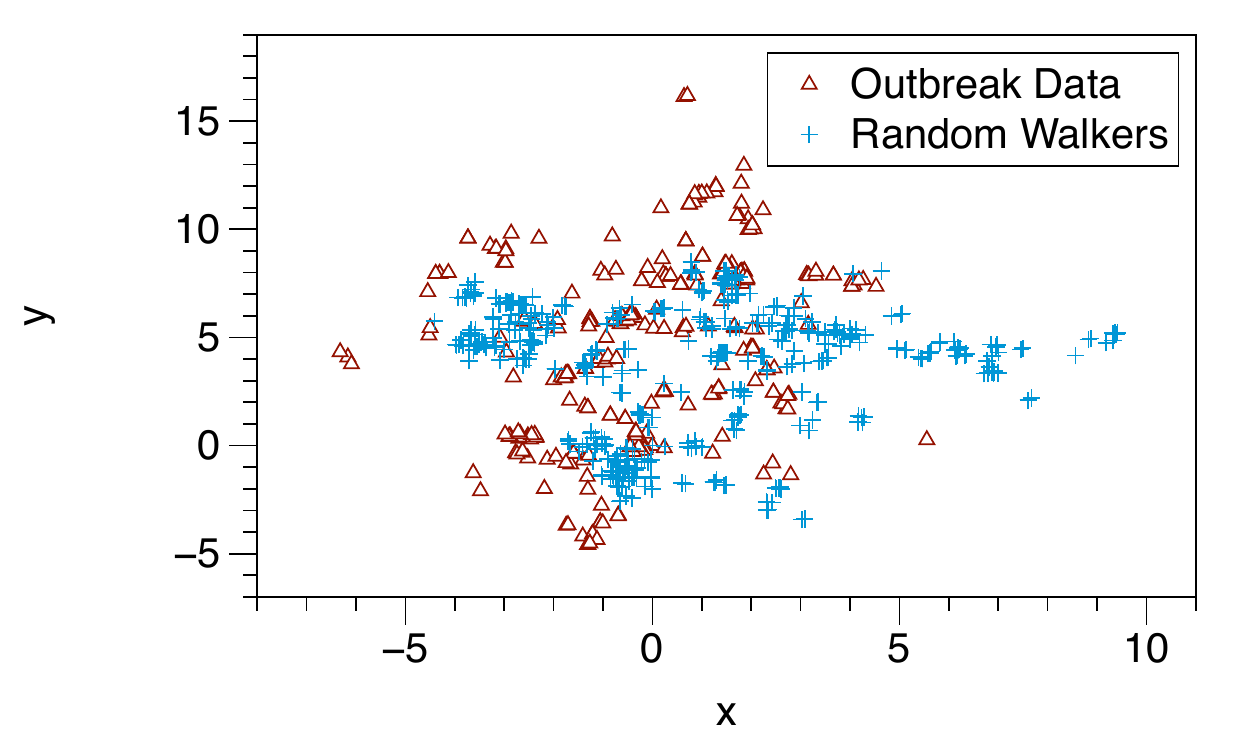}
\caption{Real outbreak data vs. one realization of a random walk for Cluster 5. The index case is set to coordinates (0, 0). Outbreak data from 121 days, random walk with multiplicity 3.0 resulting in 363 steps.
}
\label{fig:clu5_realization}
\end{center}
\end{figure}

Figure \ref{fig:clu5_realization} shows a realization of a random walk trained on the outbreak data.
The multiplicity of the process is 3.0.

\begin{figure}[htbp]
\begin{center}
\includegraphics[width=\columnwidth]{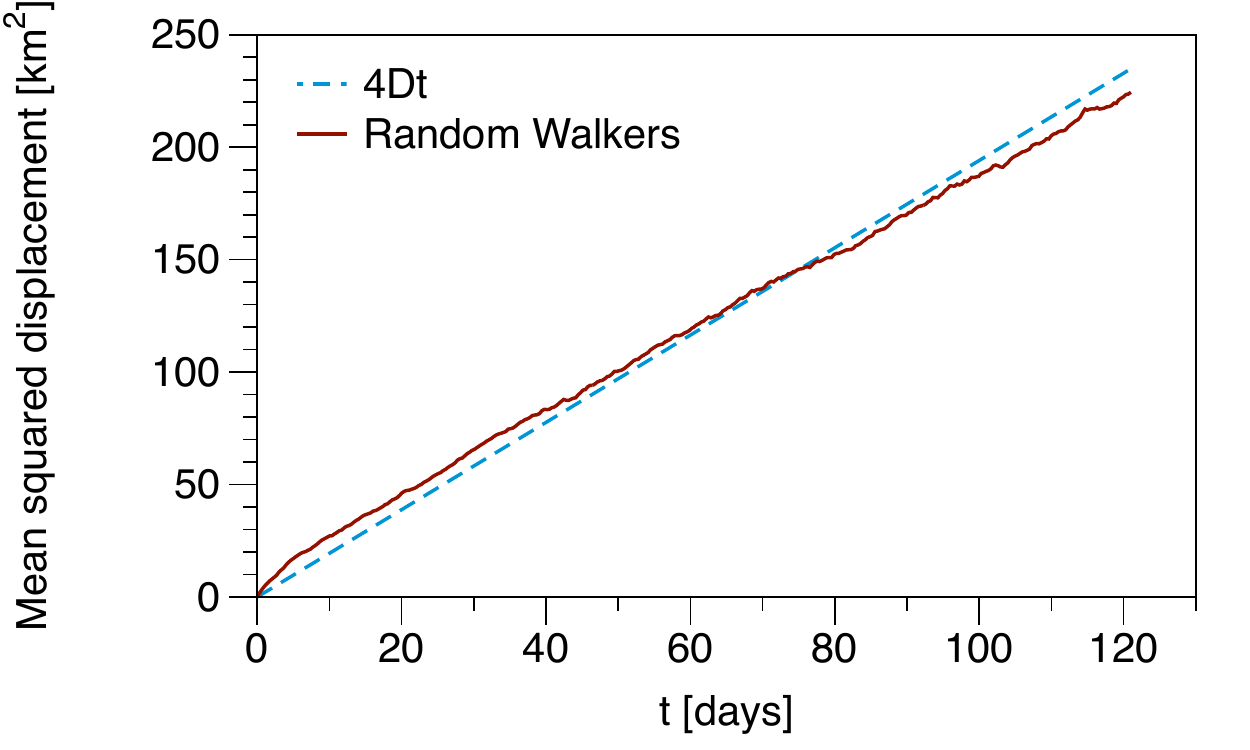}
\caption{Mean squared displacement for Cluster 5. Mean over 10,000 random walkers (red line). The resulting diffusion constant is $D = (0.16 \pm 0.01)\; km^2 / day$.}
\label{fig:clu5_msd}
\end{center}
\end{figure}

We show the mean squared displacement in Figure \ref{fig:clu5_msd}.
The diffusion coefficient is $D = (0.16 \pm 0.01) \; km^2 / day$.

\subsubsection{Cluster 6}

Figure \ref{fig:clu5_realization} shows a realization of a random walk trained on the outbreak data.
The multiplicity of the process is 6.2.
\begin{figure}[htbp]
\begin{center}
\includegraphics[width=\columnwidth]{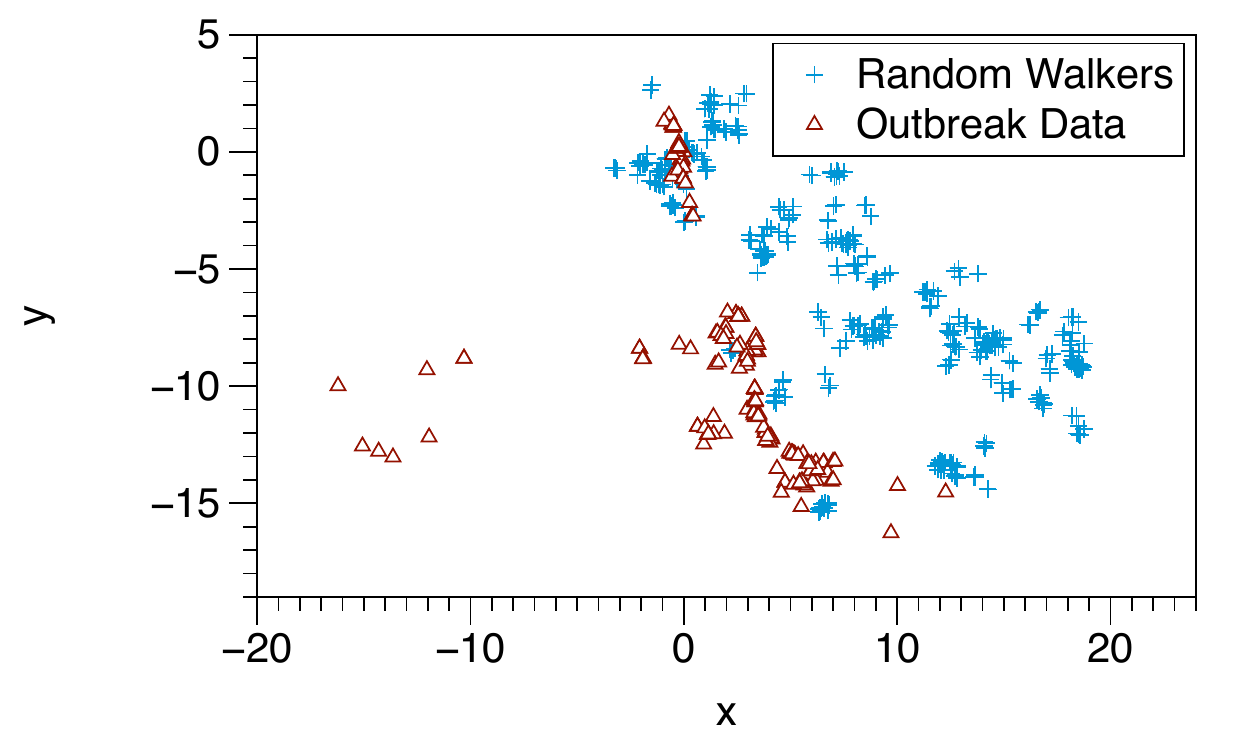}
\caption{Real outbreak data vs. one realization of a random walk for Cluster 6. The index case is set to coordinates (0, 0). Outbreak data from 127 days, random walk with multiplicity 6.2 resulting in 787 steps.
}
\label{fig:clu6_realization}
\end{center}
\end{figure}

\begin{figure}[htbp]
\begin{center}
\includegraphics[width=\columnwidth]{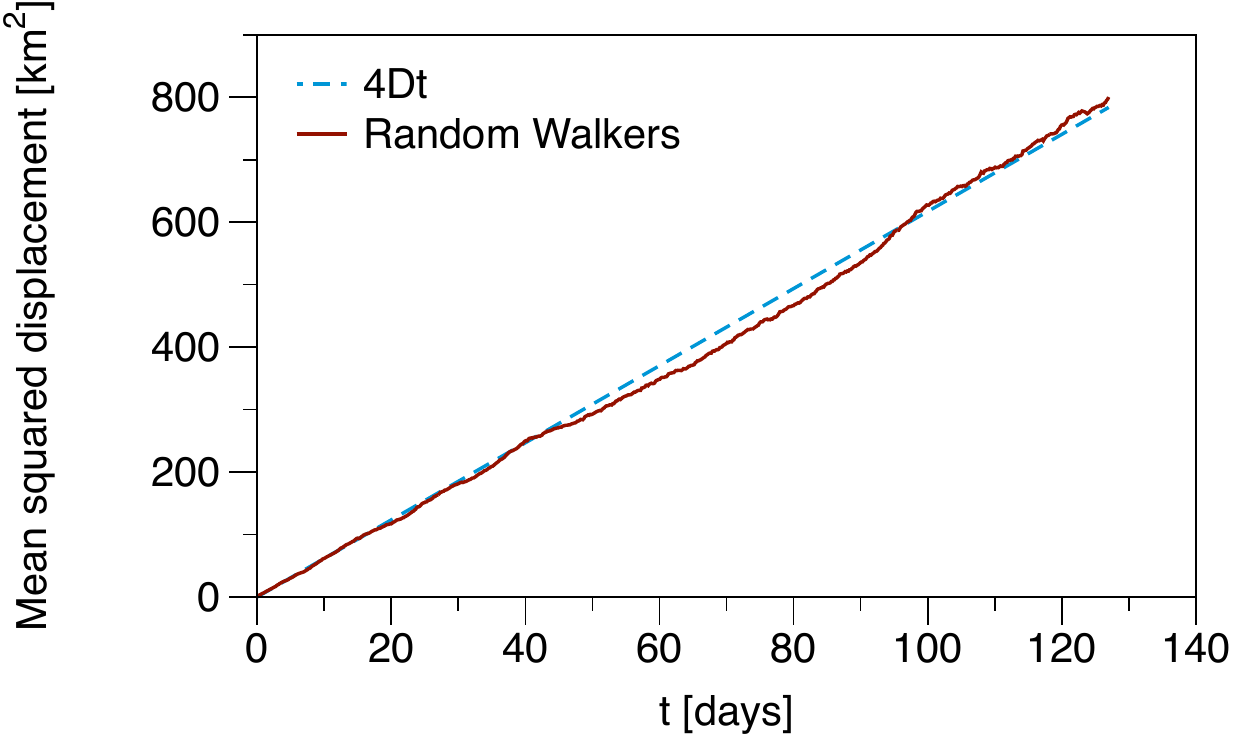}
\caption{Mean squared displacement for Cluster 6. Mean over 10,000 random walkers (red line). The resulting diffusion constant is $D = (1.54 \pm 0.07)\; km^2 / day$.}
\label{fig:clu6_msd}
\end{center}
\end{figure}

We show the mean squared displacement in Figure \ref{fig:clu6_msd}.
The diffusion coefficient is $D = (1.54 \pm 0.07) \; km^2 / day$.

\subsubsection{Summary and discussion of Clusters 2--6}
The Clusters 2, 3, and 5 show a behavior similar to Cluster 1 in the main text.
Their multiplicities are relatively low and most realizations of random walks appear very similar to the real outbreak data.

Clusters 4 and 6 show remarkable differences between synthetic and real outbreak data, as shown in Figures \ref{fig:clu4_realization} and \ref{fig:clu6_realization}.
The figures demonstrate that the generated data points cover a larger area when compared to the more compact outbreak data.
This is caused by the fact that both clusters -- and Cluster 6 in particular -- are strongly geographically constrained by data being restricted to within the German country borders.
In particularly, Cluster 6 is located along the river Oder.
This implies that a large proportion of cases on the polish side is missing in the cluster.

Since the random walk model does not take into account such geographical constraints, the random walkers move in all directions ignoring the constraints.
As a consequence, they cover a much larger area (take the eastern regions in Figure \ref{fig:clu6_realization} as an example) and thus the diffusion constant is magnified.
Cluster 4 shows a similar behavior, even if to a weaker extent (Figure \ref{fig:clu4_realization}).

An additional bias in Cluster 6 is the high multiplicity of 6.2.
This is the highest value among all clusters and it causes a strong bias in the random walk model, since the random walker has to cover more than 6 events occurring in the data each day.
However, the random walk assumption only holds for multiplicities close to 1.
Moreover, Cluster 6 is located in an urban area and consequently this restriction did not allow for the same control measures as in the other clusters.

\end{document}